\documentclass{JHEP3}
\usepackage{amssymb}
\usepackage{amsfonts}
\usepackage{mathrsfs}

\newcommand{\e}{\begin{eqnarray}}
\newcommand{\ee}{\end{eqnarray}}

\newcommand{\CN}{{\cal N}}
\newcommand{\CL}{{\cal L}}

\newcommand{\pa}{{a^{\prime}}}
\newcommand{\pb}{{b^{\prime}}}
\newcommand{\pc}{{c^{\prime}}}
\newcommand{\pd}{{d^{\prime}}}
\newcommand{\pe}{{e^{\prime}}}
\newcommand{\pf}{{f^{\prime}}}
\newcommand{\pg}{{g^{\prime}}}

\newcommand{\DA}{{\dot A}}
\newcommand{\DB}{{\dot B}}
\newcommand{\DC}{{\dot C}}
\newcommand{\DD}{{\dot D}}

\def\a{\alpha}
\def\b{\beta}
\def\d{\delta}
\newcommand{\ep}{\epsilon}
\newcommand{\g}{\gamma}
\newcommand{\p}{\psi}
\newcommand{\s}{\sigma}
\def\t{\tau}

\newcommand{\bp}{\bar{\psi}}

\title{Superspace Formulation in a Three-Algebra Approach
to $D=3, {\cal N}=4,5$ Superconformal Chern-Simons Matter
Theories}

\author{Fa-Min Chen , Yong-Shi Wu \\
Department of Physics and Astronomy, University of Utah\\
Salt Lake City, UT 84112-0830, USA \\
E-mail:
\email{fchen@physics.utah.edu}, \email{wu@physics.utah.edu}}

\abstract{We present a superspace formulation of the $D=3, {\cal
N}=4,5$ superconformal Chern-Simons Matter theories, with matter
supermultiplets valued in a symplectic 3-algebra. We first
construct an ${\cal N}=1$ superconformal action, and then
generalize a method used by Gaitto and Witten to enhance the
supersymmetry from ${\cal N}=1$ to ${\cal N}=5$. By decomposing
the ${\cal N}=5$ supermultiplets and the symplectic 3-algebra
properly and proposing a new super-potential term, we construct
the ${\cal N}$=4 superconformal Chern-Simons matter theories in
terms of two sets of generators of a (quaternion) symplectic
3-algebra. The ${\cal N}$=4 theories can also be derived by
requiring that the supersymmetry transformations are closed
on-shell. The relationship between the 3-algebras, Lie
superalgebras, Lie algebras and embedding tensors (proposed in [E. A. Bergshoeff, O. Hohm, D. Roest, H. Samtleben, and E. Sezgin, J. High Energy Phys. 09
(2008) 101.]) is also clarified. The general ${\cal
N}=4,5$ superconformal Chern-Simons matter theories in terms of
ordinary Lie algebras can be rederived in our 3-algebra approach.
All known $ {\cal N}=4,5$ superconformal Chern-Simons matter
theories can be recovered in the present superspace formulation
for super-Lie-algebra realization of symplectic 3-algebras.}


\usepackage{bbm}
\usepackage{amsmath}
\usepackage{slashed}

\begin{document}

\section{Introduction} \label{Introduction}

In the last two years, extended (${\cal N}\geq 4$) supersymmetric
Chern-Simons-matter (CSM) theories in 3D have attracted a lot of
interests in the string/$M$-theory community, because they are
natural candidates of the dual gauge theories of multi M2-branes
in $M-$theory. Less extended supersymmetric (${\cal N} < 4$) CSM
theories with arbitrary gauge groups were constructed and
investigated long time ago \cite{Zupnik:1988}-\cite{Kapustin99}.
Generic Chern-Simons gauge theories with or without (massless)
matter were demonstrated to be conformally invariant even at the
quantum level \cite{CSW1, CSW0, CSW2, Piguet, Saemann1}. However,
it was much more difficult until recently to construct ${\cal
N}\geq 4$ CSM theories, since only some special gauge groups are
allowed in these theories.

By virtue of the Nambu 3-algebra structure \cite{Nambu,Limiao},
the maximally supersymmetric ${\cal N}=8$ CSM theory with $SO(4)$
gauge group was first constructed independently by Bagger and
Lambert \cite{Bagger} and by Gustavsson \cite{Gustavsson} (BLG).
The BLG theory was conjectured to be the dual gauge theory of two
M2-branes \cite{DMPV, LambertTong, KM:May08, sissa}. The Nambu
3-algebra, equipped with a symmetric and positive-definite metric,
has the limitation that it can only generate an $SO(4)$ gauge
theory \cite{Gauntlett, Papadopoulos, MIT}, too restrictive for a
low-energy effective description of M2-branes.

Very soon Aharony, Bergman, Jafferis and Maldacena (ABJM) have
observed \cite{ABJM} that an ${\cal N}=2$ superconformal CSM theory,
with gauge group $U(N)\times U(N)$, actually has an $SU(4)$
R-symmetry, hence an enhanced supersymmetry ${\cal N}=6$. The same
theory was also obtained by taking the infrared limit of a brane
construction. In their formulation, the Nambu 3-algebra structure
did not play any role, though the ABJM theory with $SU(2)\times
SU(2)$ gauge group is equivalent to the BLG theory. Based on the
brane construction, ABJM conjectured that at level $k$ their theory
describes the low energy limit of $N$ M2-branes probing a
$\textbf{C}^4/\textbf{Z}_k$ singularity. In the special cases of
$k=1, 2$, the theory has the maximal supersymmetries (${\cal N}=8$)
\cite{ABJM, klebanov:Jun09, Gustavsson:Jun09, KOS:Jun09}. In a
large-$N$ limit the ABJM theory is then dual to $M-$theory on
$AdS_4\times S^7/\textbf{Z}_k$ \cite{ABJM}. The superspace
formulation and a manifest $SU(4)$ R-symmetry formulation of the
ABJM theory can be found in Ref. \cite{Benna} and \cite{Schwarz},
respectively.

In Ref. \cite{Bergshoeff:2008cz, Bergshoeff}, some extended
superconformal gauge theories are constructed by taking a conformal
limit of $D=3$ gauged supergravity theories. In this approach, the
embedding tensors play a crucial role.  Gaiotto and Witten (GW)
\cite{GaWi} have been able to construct a large class of ${\cal
N}=4$ CSM theories by a method that enhances ${\cal N}=1 $
supersymmetry to ${\cal N}=4$. They also demonstrated that the gauge
groups can be classified by super Lie algebras. In Ref.
\cite{HosomichiJD}, the GW theory was extended to include additional
twisted hyper-multiplets; in particular, the extended GW theory with
$SO(4)$ gauge group was demonstrated to be equivalent to the BLG
theory. In Ref. \cite{Hosomichi:2008jb}, two new theories, ${\cal
N}=5$, $Sp(2M)\times O(N)$ and ${\cal N}=6$, $Sp(2M)\times O(2)$ CSM
theories, were constructed by further enhancing the R-symmetry to
$Sp(4)$ and $SU(4)$, respectively, and the ${\cal N}=6$, $U(M)\times
U(N)$ CSM theory was rederived. The gravity duals of ${\cal N}=5$,
$Sp(2M)\times O(N)$ and ${\cal N}=6$, $U(M)\times U(N)$ theories
were studied in Ref. \cite{Aharony:2008gk}. By using group
representation theory and applying GW's super-Lie-algebra method for
classifying gauge groups, the ${\cal N}=1$ to ${\cal N}=8$ CSM
theories were constructed systematically in a recent paper
\cite{MFM:Aug09}.

The progress mentioned in the last two paragraphs was made using
mainly ordinary Lie algebras. 
On the other hand, Bagger and Lambert  have been able to
construct the ${\cal N}=6,U(M)\times U(N)$ theory in terms of a
modified 3-algebra \cite{Bagger08:3Alg}. Unlike the Nambu 3-algebra
with totally antisymmetric structure constants, the structure
constants of the modified 3-algebra are antisymmetric only in the
first two indices. By introducing an invariant antisymmetric tensor
into a 3-algebra, hence called a `symplectic 3-algebra', another
class of ${\cal N}=6$ CSM theories, with gauge group $Sp(2M)\times
O(2)$, has been constructed by the authors of the present paper
\cite{ChenWu1}. We have also demonstrated that the ${\cal
N}=6,U(M)\times U(N)$ theory can be recast into the symplectic
3-algebraic formalism \cite{ChenWu1}. In Ref. \cite{Chen2}, both the
general ${\cal N}=5$ and ${\cal N}=6$ CSM theories have been
formulated in a unified symplectic 3-algebraic framework. These
theories based on 3-algebras were constructed by requiring that the
supersymmetries must be closed on-shell. All examples of $\CN=5$ theories
were recovered in Ref. \cite{Bagger10} by specifying the 3-algebra structure constants.

One goal of the present paper is to combine the superspace
formalism with the 3-algebra, to rederive the ${\cal N}=5$
theories by using the Giatto-Witten enhancement mechanism.
Previously the ${\cal N}=5$ theories were derived from the ${\cal
N}=4$ theories by carefully choosing the gauge groups
\cite{Hosomichi:2008jb, MFM:Aug09}. So the construction of ${\cal
N}=5$ theories by enhancing ${\cal N}=1$ supersymmetry is
interesting in its own right, especially in a 3-algebraic
framework. It provides insight into the relationship between the
3-algebra and conventional Lie-algebra approach.

Another goal is to construct general ${\cal N}=4$ theories in the
(quaternion) 3-algebra framework, in which there are two similar
sets of complex 3-algebra generators. These ${\cal N}=4$ theories
are 3-algebra version of Chern-Simons quiver gauge theories. We
will construct the ${\cal N}=4$ theories by two distinct methods.
We first start from ${\cal N}=5$ supermultiplets, decompose them
and the symplectic 3-algebra generators properly, and propose a
new superpotential which is ${\cal N}$=4 superconformally
invariant. Alternatively, we will derive the same ${\cal N}$=4
theories by requiring that the supersymmetry transformations are
closed on-shell, i.e., we will examine the $\CN=4$ algebra and
check its closure. The closure of ${\cal N}=4$ algebra in the GW
theory (\emph{without} the twisted hypermultiplets) has been
checked in Ref. \cite{GaWi}. However, to our knowledge, the
closure of the algebra in theories \emph{with} the twisted
hypermultiplets has not been explicitly checked in the literature.
So our calculation will fill this gap.

We will systematically investigate the relations between the
3-algebras, Lie superalgebras, ordinary Lie algebras and embedding
tensors that are used to build $D=3$ extended supergravity theories
in Ref. \cite{Bergshoeff}. The relations between the 3-algebras and
Lie superalgebras are explored in Ref. \cite{ MFM:Aug09, Jose,
Jakob}, using representation theory. They did not discuss the
relations between the embedding tensors in Ref. \cite{Bergshoeff}
and 3-algebras or Lie superalgebras. We will fill this gap by a
more physical approach.

We will demonstrate that the symplectic 3-algebra can be realized in
terms of a super Lie algebra. The generators of the 3-algebra $T_I$
can be realized as the fermionic generators of the super Lie algebra
$Q_I$, and the 3-bracket is realized in terms of a double graded
bracket: $[T_I,T_J; T_K]\doteq [\{Q_I, Q_J\}, Q_K]$. In this
realization, the fundamental identity (FI) of the symplectic
3-algebra can be converted into the $MMQ$ Jacobi identity of the
super Lie algebra ($M$ is a bosonic generator). It will be shown
that the structure constants of the symplectic 3-algebra furnish a
quaternion representation of the bosonic part of the super Lie
algebra, and play the role of Killing-Cartan metric of the bosonic
part of the super Lie algebra. Then the FI of the 3-algebra is
rewritten as an ordinary commutator, whose structure constants are
totally antisymmetric. Moreover, we prove that the structure
constants of the symplectic 3-algebra are the components of the
embedding tensor proposed in \cite{Bergshoeff}, if we realize the
symplectic 3-algebra in terms of the super Lie algebra.

The general $\CN=4,5$ superconformal Chern-Simons-matter theories
in terms of ordinary Lie algebras can be re-drived from our
super-Lie-algebra realization of the symplectic 3-algebras. Not
only all known examples of $\CN=4,5$ ordinary CSM theories, but
also $\CN=4$ CSM quiver gauge theories (including some new
examples), can be produced as well. The details for the latter
will be presented in a forthcoming paper. Therefore, our
superspace formulation for the super-Lie-algebra realization of
symplectic 3-algebras provide a unified treatment of all known
$\CN=4,5,6,8$ CSM theories, including new examples of $\CN=4$
quiver gauge theories as well.

This paper is organized as follows. In Sec. \ref{3Alg} we
review symplectic 3-algebras and define the notations. Sec.
\ref{N1to5} is devoted to the construction of the ${\cal N}=5$
theories by enhancing the supersymmetry from ${\cal N}=1$ to
${\cal N}=5$ in a 3-algebraic framework. In Sec. \ref{secn4.1},
we derive the ${\cal N}$=4 theories by decomposing the ${\cal
N}=5$ supermultiplets and the symplectic 3-algebra properly and
proposing a new superpotential. The closure of the $\CN=4$ algebra
is explicitly verified in Sec. \ref{CloseN4}. In Sec.
\ref{TLE}, we discuss the relations between 3-algebras, super Lie
algebras, ordinary Lie algebras and the embedding tensors proposed
in Ref. \cite{Bergshoeff}. In Sec. \ref{LieN4N5}, we present
how to reproduce the Lie algebra version of $\CN=4, 5$ theories
from the 3-algebra approach. The last Sec. is devoted to
conclusions.


\section{${\cal N}=5$ theories and Symplectic Three-Algebras}\label{N5}
\subsection{A Review of Symplectic Three-Algebra}\label{3Alg}

A 3-algebra is a complex vector space equipped a 3-bracket, mapping
three vectors to one vector \cite{Chen2}:
\begin{eqnarray}\label{Symp3Bracket}
[T_I,T_J;T_K]=f_{IJK }{}^LT_{L},
\end{eqnarray}
where $T_I$ ($I=1,2,\ldots,M$) is a set of generators. The set of
complex numbers $f_{IJK}{}^L$ are called the structure constants. We
define the global transformation of a field $X$ valued in this
3-algebra ($X=X^KT_K$) as \cite{Bagger}:
\begin{eqnarray}\label{GlbTran}
\delta_{\tilde\Lambda}X=\Lambda^{IJ}[T_I,T_J;X],
\end{eqnarray}
where the parameter $\Lambda^{IJ}$ is independent of spacetime
coordinate. (The symmetry transformation (\ref{GlbTran}) will be
gauged later). For (\ref{GlbTran}) to a symmetry, one has to require
that it acts as a derivative \cite{Bagger}:
\begin{eqnarray}
\delta_{\tilde\Lambda}([X,Y;Z])=
[\delta_{\tilde\Lambda}X,Y;Z]+[X,\delta_{\tilde\Lambda}Y;Z]
+[X,Y;\delta_{\tilde\Lambda}Z],
\end{eqnarray}
where $Y=Y^NT_N$ and $Z=Z^KT_K$. Canceling $\Lambda^{IJ}, X^M, Y^N$
and $Z^K$ from both sides, we obtain the following FI satisfied by the generators:
\begin{equation}\label{FI}
[T_I,T_J; [T_M,T_N;T_K]]=[[T_I,T_J;T_M],T_N;
T_K]+[T_M,[T_I,T_J;T_N]; T_K]+[T_M,T_N; [T_I,T_J;T_K]].
\end{equation}
The FI is a generalization of the Jacobi identity of an ordinary Lie
algebra. Combining the three-bracket (\ref{Symp3Bracket}) and the FI
(\ref{FI}), we find that the FI satisfied by the structure constants
is
\begin{equation}\label{FFI}
f_{MNK}{}^Of_{IJO}{}^{L}=f_{IJM}{}^Of_{ONK}{}^{L}
+f_{IJN}{}^Of_{MOK}{}^{L}+f_{IJK}{}^Of_{MNO}{}^{L}.
\end{equation}

To define a symplectic 3-algebra, we introduce a symplectic bilinear
form into the 3-algebra:
\begin{equation}\label{SympInnerProdu}
\omega(X,Y)=\omega_{IJ}X^IY^J.
\end{equation}
We denote the inverse of the antisymmetric tensor $\omega_{IJ}$
as $\omega^{IJ}$. The existence of the inverse implies that a
3-algebra index $I$ must run from $1$ to $M=2L$. We will use
$\omega_{IJ}$ and $\omega^{IJ}$ to lower or raise 3-algebra indices;
for instance, $f_{IJKL}\equiv\omega_{LM}f_{IJK}{}^{M}$. The
symplectic bilinear form must be invariant under an arbitrary global
transformation:
\begin{eqnarray}\label{DeltaOmg}
\delta_{\tilde\Lambda}(\omega_{IJ}X^IY^J)
&=&\Lambda^{LM}(f_{LMI}{}^K\omega_{KJ}
+f_{LMJ}{}^K\omega_{IK})X^IY^J\nonumber\\
&=&0.
\end{eqnarray}
It turns out that the structure constants must be \emph{symmetric}
in the last two indices:
\begin{eqnarray}\label{SymInLast2Ind}
f_{LMIJ}=f_{LMJI}.
\end{eqnarray}
From point of view of ordinary Lie group, the infinitesimal matrices
\begin{equation}\tilde{\Lambda}^K{}_I\equiv
\Lambda^{LM}f_{LM}{}^K{}_I\end{equation} must form the Lie algebra
$Sp(2L,\mathbb{C})$. We call the 3-algebra defined by the above
equations a symplectic 3-algebra.

Since the 3-algebra is also a complex vector space, one can define a
Hermitian bilinear form
\begin{eqnarray}\label{HermiInnerProd}
h(X,Y)=X^{*I}Y^{I}
\end{eqnarray}
(with $X^{*I}$ the complex conjugate of $X^I$), which is naturally
positive-definite and will be used to construct the Lagrangians. The
Hermitian bilinear form is also required to be invariant under the
global transformation:
\begin{eqnarray}\label{InvOfHermInner}
\delta_{\tilde\Lambda}(X^{*I}Y^I)
&=&(\Lambda^{*LM}f^{*}_{LMI}{}^K+\Lambda^{LM}f_{LMK}{}^I)
X^{*I}Y^K\nonumber\\
&=&0.
\end{eqnarray}
To solve the above equation, we assume that the parameter
$\Lambda^{LM}$ is Hermitian: $\Lambda^{*LM}=\Lambda_{ML}$. Since it
also carries two symplectic 3-algebra indices, it obeys the natural
reality condition
$\Lambda^{*LM}=\omega_{LI}\omega_{MJ}\Lambda^{IJ}$. These two
equations imply that the parameter is symmetric, i.e.
$\Lambda_{ML}=\Lambda_{LM}$. In summary, we have
\begin{equation}\label{parameter}
\Lambda^{*LM}=\Lambda_{ML}=\Lambda_{LM}.
\end{equation}
Now since the parameter $\Lambda^{IJ}$ is symmetric, re-examining
the global transformation (\ref{GlbTran}) leads us to require that
the structure constants are symmetric in the first two indices:
\begin{eqnarray}{\label{SymIn1stInd}}
f_{IJKL}=f_{JIKL}.
\end{eqnarray}
With Eq. (\ref{parameter}) and (\ref{SymIn1stInd}), we find that Eq.
(\ref{InvOfHermInner}) can be satisfied if we impose the following
reality condition on the structure constants:
\begin{eqnarray}\label{HermiCondiOnF}
f^*_{LMIK}=f^{MLKI}\quad {\rm or}\quad
f^{*L}{}_M{}^I{}_K=f^{M}{}_L{}^K{}_I.
\end{eqnarray}
Now both the symplectic bilinear form (\ref{SympInnerProdu}) and the
Hermitian bilinear (\ref{HermiInnerProd}) form are invariant under
the global transformation (\ref{GlbTran}). So from point of view of
ordinary Lie group, the symmetry group generated by the 3-algebra
transformations (\ref{GlbTran}) is nothing but $Sp(2L)$, which is
the intersection of $U(2L)$ and $Sp(2L, \mathbb{C})$.

Later we will see, to enhance the super-symmetry from ${\cal N}=1$
to ${\cal N}=5$, we will require the 3-bracket to satisfy an
additional constraint condition:
\begin{eqnarray}\label{ConstraintOn3Bracket}
\omega([T_I,T_{(J};T_K],T_{L)})=0,
\end{eqnarray}
or simply $f_{I(JKL)}=0$. Combining Eq. (\ref{ConstraintOn3Bracket})
with (\ref{SymInLast2Ind}) and (\ref{SymIn1stInd}), we have that
$f_{(IJK)L}=0$ and $f_{IJKL}=f_{KLIJ}$. In summary, the structure
constants $f_{IJKL}$ enjoy the symmetry properties
\begin{eqnarray}\label{SymmeOfF}
f_{IJKL}=f_{JIKL}=f_{JILK}=f_{KLIJ}.
\end{eqnarray}

\subsection{${\cal N}=5$ Theories in Terms of 3-Algebras}\label{N1to5}
In this subsection, we will generalize Giaotto and Witten's idea and
method \cite{GaWi} to enhance the super-symmetry from ${\cal N}=1$
to ${\cal N}=5$. \footnote{In Ref. \cite{Mauri}, the $\CN=8$ BLG theory was constructed by using $\CN=1$ superspace formulation in the Nambu 3-algebra approach.} We will work in a three-algebraic framework.

Let us first explain the mechanism for supersymmery
enhancement. We assume that the ${\cal N}=1$ superfields for the
matter fields are 3-algebra valued (our notation and convention are
summarized in appendix \ref{Identities}):
\begin{equation}\label{hypermlti}
\Phi^I_A=Z^I_A+i\theta\gamma_A{}^B\psi^I_B-\frac{i}{2}\theta^2F^I_A,
\end{equation}
where $I$ is a 3-algebra index, $A,B$ are $Sp(4)\cong SO(5)$ indices
($A,B=1,..., 4$), and $\gamma_A{}^B$ is a Hermitian $SO(5)\equiv
Sp(4)$ gamma matrix, satisfying
$\gamma_A{}^B\gamma_B{}^C=\delta_A{}^C$. \footnote{Generally
$\gamma_A{}^B\equiv c_m\gamma^{m}_A{}^B$ ($m=1,...,5$), where
$\gamma^{m}_A{}^B$ are the $SO(5)$ gamma matrices (see appendix
\ref{SO5}), and $c_m$ real coefficients. We normalize the parameters
$c_m$ so that $\delta^{mn}c_mc_n=1$. The non-uniqueness of this
gamma matrix is exactly what are allowed by the R-symmetry $SO(5)$.}
The superfield $\Phi$ satisfies the reality condition:
\begin{equation}\label{rltc}
\bar{\Phi}^A_I=\Phi^{\dag
I}_A=\omega^{AB}\omega_{IJ}\Phi^J_B.
\end{equation}
The purpose for introducing the gamma matrix into the second term
of (\ref{hypermlti}) is the following: after we promote the
supersymmetry from $\CN=1$ to $\CN=5$, we want the supercharges
and the matter fields to transform as the \textbf{5} and
\textbf{4} of $Sp(4)$, respectively, with the gamma matrix being
the couplings.

Despite that $\Phi^I_A$ carries an $Sp(4)$ index, it is still an
${\cal N}=1$ superfields in that it just depends on one copy of
fermionic coordinates $\theta^\alpha$. Generally speaking, if we
use $(\ref{hypermlti})$ to construct an ${\cal N}=1$ CSM theory,
the Yukawa couplings will contain the gamma matrix $\gamma_A{}^B$,
which is not $Sp(4)$ invariant. \footnote{With the standard
definition
$\Sigma_A{}^B\equiv\frac{1}{2}\omega_{mn}\Sigma^{mnB}_{A}$, where
$\Sigma^{mn}=\frac{1}{4}[\gamma^m, \gamma^n]$, we note that
\begin{equation}
\delta\gamma_A{}^B\equiv\Sigma_A{}^C\gamma_C{}^B-\Sigma_C{}^B\gamma_A{}^C
=\omega_{mn}c^n\gamma_A^{mB}\nonumber.
\end{equation}
Thus, $\gamma_A{}^B$ is \emph{not} $Sp(4)$ invariant.} As a
result, the CSM theory is generally not $Sp(4)$ invariant.
However, we are be able to remove the gamma matrix $\gamma_A{}^B$
from the theory by adjusting the superspace couplings. The
resulting theory then have an $Sp(4)$ global symmetry, which does
\emph{not} commute with the ${\cal N}=1$ supersymmetry. Namely the
supercharge transforms non-trivially under the $Sp(4)$ global
symmetry group. More precisely, the supercharges transform in the
vector representation of $SO(5)$ or \textbf{5} of $Sp(4)$. As a
result, the supersymmetry gets enhanced from ${\cal N}=1$ to
${\cal N}=5$. We will explain this point in details when we
examine the supersymmetry transformations.

To construct the ${\cal N}=1$ CSM theory, we first gauge the global
symmetry transformation (\ref{GlbTran}). We define the gauge
transformation of the superfield $\Phi^I$ as
\begin{equation}
\delta_{\tilde{\Lambda}}\Phi^I_A=\Lambda^{KL}f_{KL}{}^I{}_J\Phi^J_A
=\tilde{\Lambda}^I{}_J\Phi^J_A,
\end{equation}
where the parameter $\Lambda^{KL}$ is a superfield, depending on the
coordinates of the superspace. We then define the covariant
derivatives as
\begin{eqnarray}
(D_\alpha)^I{}_J=\mathscr{D}_\alpha\delta^I{}_J
+\tilde{\Gamma}_{\alpha}{}^I{}_J
\quad {\rm and} \quad
(D_\mu)^I{}_J=\partial_\mu\delta^I{}_J+\tilde{\Gamma}_{\mu}{}^I{}_J,
\end{eqnarray}
where $\mathscr{D}_\alpha$ is the super-covariant derivative,
defined by Eq. (\ref{SCD}). In accordance with our basic definition
(\ref{GlbTran}), it is natural to assume that the super-connections
take the following forms
\begin{equation}\label{Sconn}
\tilde{\Gamma}_{\alpha}{}^I{}_J\equiv\Gamma^{KL}_\alpha
f_{KL}{}^I{}_J \quad {\rm and}\quad
\tilde{\Gamma}_{\mu}{}^I{}_J\equiv\Gamma^{KL}_\mu f_{KL}{}^I{}_J,
\end{equation}
transforming as \footnote{In this section, we define a general tilde
field $\tilde{\Psi}$ as
$\tilde{\Psi}^I{}_J\equiv\Psi^{KL}f_{KL}{}^I{}_J$, where $\Psi^{KL}$
can be a superfield or an ordinary field.}
\begin{equation}\delta_{\tilde{\Lambda}}\tilde{\Gamma}_{\alpha}{}^I{}_J
=-D_\alpha\tilde{\Lambda}^I{}_J\quad \rm{and}\quad
\delta_{\tilde{\Lambda}}\tilde{\Gamma}_{\mu}{}^I{}_J
=-D_\mu\tilde{\Lambda}^I{}_J, \end{equation}
 respectively. In the
Wess-Zumino gauge, the super-connection $\tilde{\Gamma}_\alpha$
takes the form
\begin{eqnarray}\label{vecsp}\nonumber
\tilde{\Gamma}_\alpha{}^I{}_J
&=&i\theta^\beta\tilde{A}_{\alpha\beta}{}^I{}_J
+\theta^2\tilde{\chi}_{\alpha}{}^I{}_J\\
&=&(i\theta^\beta A_{\alpha\beta}^{KL}
+\theta^2\chi_{\alpha}^{KL})f_{KL}{}^I{}_J,
\end{eqnarray}
where $\tilde{\chi}_{\alpha}{}^I{}_J$ is superpartner of the gauge
field $\tilde{A}_{\alpha\beta}{}^I{}_J$. In accordance with
$(\ref{parameter})$, we assume that $A_{\alpha\beta}^{KL}$ and
$\chi_{\alpha}^{KL}$ are Hermitian and symmetric in $KL$. The two
superconnections (\ref{Sconn}) should not be independent, since
there is only one gauge symmetry. Actually, imposing the
conventional constraint \cite{Gates}
\begin{equation}\label{SComu}
\{D_\alpha, D_\beta\}=2iD_{\alpha\beta}
\end{equation}
determines the vector superconnection:
\begin{equation}\label{VSconn}
\tilde{\Gamma}_{\alpha\beta}{}^I{}_J=\tilde{A}_{\alpha\beta}{}^I{}_J-
i\theta_\alpha\tilde{\chi}_{\beta}{}^I{}_J
-i\theta_\beta\tilde{\chi}_{\alpha}{}^I{}_J+ \frac{i}
{2}\theta^2\tilde{F}_{\alpha\beta}{}^I{}_J,
\end{equation}
where the field strength is defined as
\begin{equation}
\tilde{F}_{\alpha\beta}{}^I{}_J=\frac{1}{2}(\partial_\alpha{}^\gamma
\tilde{A}_{\gamma\beta}{}^I{}_J+\partial_\beta{}^\gamma
\tilde{A}_{\gamma\alpha}{}^I{}_J)+\frac{1}{2}[\tilde{A}_{\alpha}{}^\gamma
,\tilde{A}_{\gamma\beta}]^I{}_J;\quad \tilde{F}_{\mu\nu}{}^I{}_J
=\frac{1}{2}(\gamma_{\mu\nu})^{\alpha\beta}\tilde{F}_{\alpha\beta}{}^I{}_J.
\end{equation}
The superfield $
\Gamma^{KL}_\mu=-\frac{1}{2}\g_\mu^{\a\b}\Gamma_{\alpha\beta}^{KL}$
in Eq. (\ref{Sconn}) can be read off from Eq. (\ref{VSconn}) by
re-writing the field strength as a product of a field and the
structure constants:
\begin{eqnarray}\label{Strength}\nonumber
\tilde{F}_{\alpha\beta}{}^I{}_J&=&\frac{1}{2}[\partial_{\alpha}{}^\gamma
A^{KL}_{\gamma\beta}+\partial_{\beta}{}^\gamma A^{KL}_{\gamma\alpha}+
(\tilde{A}_\alpha{}^\gamma)^L{}_M A^{MK}_{\gamma\beta}+(\tilde{A}_\beta{}^\gamma)^K{}_M
A^{ML}_{\gamma\alpha}]f_{KL}{}^I{}_J\\
&\equiv&F_{\alpha\beta}^{KL}f_{KL}{}^I{}_J.
\end{eqnarray}
In the first line we have used the FI (\ref{FFI}).

To be self-consistent, the covariant derivative $D_\alpha$ must
satisfy the Jacobi identity:
\begin{equation}\label{SJac}
[D_\alpha, \{D_\beta, D_\gamma\}]+[D_\beta, \{D_\gamma,
D_\alpha\}]+[D_\gamma, \{D_\alpha, D_\beta\}]=0.
\end{equation}
 The Jacobi identity can be solved by introducing a superfield strength $\tilde{\mathcal{W}}_{\alpha}$ \cite{Gates}:
\begin{equation}
[D_\alpha,
D_{\beta\gamma}]=i\epsilon_{\alpha\beta}\tilde{\mathcal{W}}_{\gamma}
+i\epsilon_{\alpha\gamma}\tilde{\mathcal{W}}_{\beta}.
\end{equation}
By direct calculation, we obtain
\begin{eqnarray}\label{SStrength}\nonumber
\tilde{\mathcal{W}}_{\alpha}{}^I{}_J&=&\tilde{\chi}_{\alpha}{}^I{}_J+
\theta^\beta\tilde{F}_{\alpha\beta}{}^I{}_J-\frac{i}{2}\theta^2(D_\alpha{}
^\beta\tilde{\chi}_{\beta}){}^I{}_J\nonumber\\
&=&[\chi_{\alpha}^{KL}+ \theta^\beta
F_{\alpha\beta}^{KL}-\frac{i}{2}\theta^2(D_\alpha{}
^\beta\chi_{\beta})^{KL}]f_{KL}{}^I{}_J\nonumber\\
&\equiv&\mathcal{W}_{\alpha}^{KL}f_{KL}{}^I{}_J,
\end{eqnarray}
with
\begin{equation}
(D_\alpha{} ^\beta\chi_{\beta})^{KL}f_{KL}{}^I{}_J\equiv
[\partial_{\alpha}{}^\beta
\chi^{KL}_{\beta}+(\tilde{A}_\alpha{}^\beta)^L{}_M
\chi^{MK}_{\beta}+(\tilde{A}_\alpha{}^\beta)^K{}_M
\chi^{MJ}_{\beta}]f_{KL}{}^I{}_J.
\end{equation}
In deriving the above equation, we have used the FI (\ref{FFI})
again. Here we would like to make one comment on the relation
between the FI (\ref{FFI}) and the anti-commutator (\ref{SComu}) and
the Jacobi identity (\ref{SJac}). Without consulting the FI, one
would not be able to derive Eq. (\ref{Strength}) and write
$\tilde{\Gamma}_{\alpha\beta}{}^I{}_J$ as
$\Gamma_{\alpha\beta}^{KL}f_{KL}{}^I{}_J$. This would be
inconsistent with our assumption (\ref{Sconn}) or the basic
definition (\ref{GlbTran}). Similarly, the superfield strength would
not take the form
$\tilde{\mathcal{W}}_{\alpha}{}^I{}_J=\mathcal{W}_{\alpha}^{KL}f_{KL}{}^I{}_J$
without the FI (see Eq. (\ref{SStrength})). Recall that the vector
superconnection and the superfield strength are defined through
(\ref{SComu}) and (\ref{SJac}), respectively. So, had we not
introduce the FI in Sec. \ref{3Alg}, we would have to introduce
the FI in this subsection for making the 3-bracket (\ref{GlbTran})
consistent with (\ref{SComu}) and (\ref{SJac}).

After gauging the symmetry (\ref{GlbTran}) in the superspace, we
are ready to construct an ${\cal N}=1$ CSM theory. A general
${\cal N}=1$ CSM theory consists of three parts: ${\cal L}={\cal
L}_{{\rm kin}} + {\cal L}_{{\rm CS}} + {\cal L}_W$, where ${\cal
L}_{{\rm kin}}$ is the Lagrangian of the kinetic terms of the
matter fields, ${\cal L}_{{\rm CS}}$ the Chern-Simons term and
${\cal L}_W$ the superpotential. The first part ${\cal L}_{{\rm
kin}}$ is standard:
\begin{eqnarray}\label{Lkin}
{\cal L}_{{\rm kin}}&=&\frac{1}{8}\int d^2\theta
D^\alpha\bar{\Phi}^A_ID_\alpha\Phi^I_A\\ \nonumber
&=&\frac{1}{2}(-D_\mu\bar{Z}^A_ID^\mu
Z^I_A+i\bar{\psi}^A_I\gamma_\mu D^\mu\psi^I_A+2if_{IJKL}
\gamma_B{}^A\bar{\psi}^{BK}\chi^{IJ}Z^L_A +\bar{F}^A_IF^I_A).
\end{eqnarray}
The covariant derivatives are given by
\begin{eqnarray}
D_\mu Z^A_I &=&
\partial_\mu Z^A_I -\tilde A_\mu{}^J{}_IZ^A_J,\\
D_\mu Z_A^I &=&
\partial_\mu Z_A^I +\tilde A_\mu{}^I{}_JZ_A^J.
\end{eqnarray}
We propose the Chern-Simons term as
\begin{eqnarray}\label{LCS}
{\cal L}_{{\rm CS}}&=&\frac{1}{8}\int
d^2\theta[-if_{IJKL}\Gamma^{\alpha IJ}
\mathcal{W}_{\alpha}^{KL}+\frac{1}{3}f_{IJK}{}^Of_{OLMN}\Gamma^{\alpha
IJ}\Gamma^{\beta KL}\Gamma^{MN}_{\alpha\beta}]\\ \nonumber
&=&\frac{1}{2}\epsilon^{\mu\nu\lambda}(f_{IJKL}A_\mu^{IJ}\partial_\nu
A_\lambda^{KL}+\frac{2}{3}f_{IJK}{}^Of_{OLMN}A_\mu^{IJ}A_\nu^{KL}A_\lambda^{MN})+
\frac{i}{2}f_{IJKL}\chi^{\alpha IJ}\chi_\alpha^{KL}.
\end{eqnarray}
The first part of the second line is precisely the `twisted'
Chern-Simons term in Ref. \cite{Chen2}, while the gaugino $\chi$ is
just an auxiliary field, whose equation of motion is
\begin{equation}\label{Chi}
\chi^{\alpha IJ}=-\gamma_B{}^A\psi^{\alpha B(I}Z^{J)}_A.
\end{equation}
Substituting it into (\ref{Lkin}) and (\ref{LCS}) gives a Yukawa
coupling:
\begin{equation}\label{YKW1}
-\frac{i}{2}Z^I_AZ^J_B\psi^K_C\psi^L_Df_{IKJL}\gamma^{AC}\gamma^{BD}.
\end{equation}
Note that this term is not $Sp(4)$ invariant, because the gamma
matrix is \emph{not} $Sp(4)$ invariant (see footnote 2).

Let us now consider the superpotential $W(\Phi)$. It must satisfy
two conditions. First, for conformal invariance, the superpotential
must be homogeneous and quartic in $\Phi$; schematically,
$W(\Phi)\sim\Phi\Phi\Phi\Phi$. Second, after combining
(\ref{YKW1}) with the Yukawa terms arising from $W(\Phi)$, the final
expression must be $Sp(4)$ invariant. Before proposing $W(\Phi)$, it
is useful to look at the structure of (\ref{YKW1}): it contains
$\gamma^{AC}\gamma^{BD}$. The essential observation is that
$\gamma^{[AC}\gamma^{BD]}$ has to be proportional to the totally
antisymmetric (invariant) tensor $\varepsilon^{ABCD}$, since this
tensor is unique in $Sp(4)$. The precise expression is
\begin{eqnarray}\label{Sp4ID}
-\varepsilon^{ABCD}
&=&\gamma^{AC}\gamma^{BD}-\gamma^{BC}\gamma^{AD}+\gamma^{BA}\gamma^{CD}\\
&=&\omega^{AB}\omega^{CD}-\omega^{AC}\omega^{BD}+\omega^{AD}\omega^{BC}\nonumber.
\end{eqnarray}
Namely, our problem may be solved if the final expression for
(\ref{YKW1}) plus the Yukawa terms arising from $W(\Phi)$ is
somehow related to (\ref{Sp4ID}). So we are inspired to propose
the following superpotential
\begin{eqnarray}\label{SuperPot0} W(\Phi)=\frac{1}{4}(g_{IJKL}\omega^{AB}
\omega^{CD}\Phi^{I}_A\Phi^{J}_B\Phi^{K}_C\Phi^{L}_D+
\tilde{g}_{IJKL}\gamma^{AB}\gamma^{CD}\Phi^{I}_A\Phi^{J}_B\Phi^{K}_C\Phi^{L}_D),
\end{eqnarray}
where the 3-algebra tensor $g$ satisfies
$g_{IJKL}=-g_{JIKL}=-g_{IJLK}=g_{KLIJ}$, and $\tilde{g}$ has the
same symmetry properties. We require that the tensors $g$ and
$\tilde{g}$ are gauge invariant. This implies that $g$ and
$\tilde{g}$ can be expressed in terms of $\omega_{IJ}$ and
$f_{IJKL}$, the only two gauge invariant quantities. After carrying
out the Berezin integration $\frac{i}{2}\int d^2\theta W(\Phi)$, we
obtain
\begin{eqnarray}\label{SuperPot}
{\cal L}_W
&=&-\frac{i}{2}Z^I_AZ^J_B\psi^K_C\psi^L_D(g_{IJKL}\omega^{AB}\omega^{CD}
+2g_{IKJL}\gamma^{AC}\gamma^{BD}+\tilde{g}_{IJKL}\gamma^{AB}\gamma^{CD}
+2\tilde{g}_{IKJL}\omega^{AC}\omega^{BD})\nonumber\\
&&-(g_{IJKL}\omega^{AB}\omega^{CD}+\tilde{g}_{IJKL}\gamma^{AB}
\gamma^{CD})Z^J_BZ^K_CZ^L_DF^I_A.
\end{eqnarray}
The first and last term of the first line are already $Sp(4)$
invariant. Combining the middle two terms of the first line with
(\ref{YKW1}) gives
\begin{eqnarray}\label{YKWT}
&&-\frac{i}{2}Z^I_AZ^J_B\psi^K_C\psi^L_D[(2g_{IKJL}+f_{IKJL})\gamma^{AC}\gamma^{BD}+
\tilde{g}_{IJKL}\gamma^{AB}\gamma^{CD}].
\end{eqnarray}
Since we wish to use Eq. (\ref{Sp4ID}), we first have to
anti-symmetrize $AB$ in the expression $\gamma^{AC}\gamma^{BD}$.
Equivalently, we have to set the part proportional to
$Z^{(I}_{(A}Z^{J)}_{B)}$ to be zero:
\begin{eqnarray}\label{g}
g_{IKJL}+g_{JKIL}+\frac{1}{2}f_{IKJL}+\frac{1}{2}f_{JKIL}=0.
\end{eqnarray}
Now the remaining part of (\ref{YKWT}) is antisymmetric in $AB$:
\begin{eqnarray}\label{YKWT3}
&&\frac{i}{2}Z^I_AZ^J_B\psi^K_C\psi^L_D[(2g_{IKJL}+f_{IKJL})
\gamma^{C[A}\gamma^{B]D}-\tilde{g}_{IJKL}\gamma^{AB}\gamma^{CD}].
\end{eqnarray}
It can be seen that if we set
\begin{eqnarray}\label{gtld}
\tilde{g}_{IJKL}=-\frac{1}{2}(g_{IKJL}-g_{JKIL}+\frac{1}{2}f_{IKJL}-\frac{1}{2}f_{JKIL})
\end{eqnarray}
and apply the key identity (\ref{Sp4ID}), then Eq. (\ref{YKWT3})
becomes
\begin{eqnarray}\label{YKWT2}
\frac{i}{2}Z^I_AZ^J_B\psi^K_C\psi^L_D\tilde{g}_{IJKL}
(\omega^{AB}\omega^{CD}-\omega^{AC}\omega^{BD}+\omega^{AD}\omega^{BC}).
\end{eqnarray}
Now Eq. (\ref{YKWT2}) is manifestly $Sp(4)$ invariant. However we
still need to solve (\ref{g}) and (\ref{gtld}) in terms of
$f_{IJKL}$ and $\omega_{IJ}$. An equation similar to (\ref{g}) is
first derived by GW \cite{GaWi}:
\begin{eqnarray}\label{g2}
g_{IKJL}+g_{JKIL}+\frac{3}{4}k_{mn}\tau^m_{IK}\tau^n_{JL}+\frac{3}{4}k_{mn}\tau^m_{JK}\tau^n_{IL}=0,
\end{eqnarray}
where the set of matrices $\tau^m_{IK}$ is in the fundamental
representation of $Sp(2L)$ or its subalgebra, and $k_{mn}$ is the
Killing-Cartan metric. Although the (${\cal N}=4$) GW theory is not
an ${\cal N}=5$ theory, the similarity between (\ref{g}) and
(\ref{g2}) strongly suggests that $f_{IJKL}$ can be specified as
$k_{mn}\tau^m_{IJ}\tau^n_{KL}$ (up to an unimportant constant). This
is indeed the case: the FI (\ref{FFI}) does admit an explicit
solution in terms of the tensor product
$f_{IJKL}=k_{mn}\tau^m_{IJ}\tau^n_{KL}$. It is straightforward to
verify that $f_{IJKL}=k_{mn}\tau^m_{IJ}\tau^n_{KL}$ satisfy the FI
(\ref{FFI}). This solution is first found by Gustavsson by
converting the FI into two independent commutators of ordinary Lie
algebra \cite{Gustavsson}. Later we will discuss the relations
between the 3-algebra and the ordinary Lie algebra in details. Eq.
(\ref{g}) can be easily solved by adopting a method in Ref.
{\cite{GaWi}}. Summing (\ref{g}) over cyclic permutations of $IKJ$
gives
\begin{equation}\label{Constr2}
f_{(IKJ)L}=0,\quad {\rm or} \quad f_{I(KJL)}=0.
\end{equation}
This is precisely (\ref{ConstraintOn3Bracket}), as we stated
earlier. The above equation is also derived by requiring that the
${\cal N}=5$ supersymmetry transformations are closed on-shell
\cite{Chen2}. Eq. (\ref{g}) is solved by setting
\begin{equation}\label{g3}
g_{IKJL}=\frac{1}{6}(f_{IJKL}-f_{ILKJ}).
\end{equation}
Substituting (\ref{g3}) into (\ref{gtld}), we obtain
\begin{equation}\label{gtld2}
\tilde{g}_{IJKL}=\frac{1}{3}(f_{ILJK}-f_{IKJL}).
\end{equation}
Substituting (\ref{gtld2}) into (\ref{YKWT2}), then combining
(\ref{YKWT2}) with the first and the last term of the first line of
(\ref{SuperPot}), we reach the final expression for all Yukawa
terms:
\begin{equation}
-\frac{i}{2}\omega^{AB}\omega^{CD}f_{IJKL}(Z^I_AZ^K_B\psi^J_C\psi^L_D-
2Z^I_AZ^K_D\psi^J_C\psi^L_B).
\end{equation}

Finally we integrate out the auxiliary field $F^I_A$ appearing in
(\ref{Lkin}) and (\ref{SuperPot}):
\begin{eqnarray}\label{F}
\bar{F}^A_I=\frac{1}{3}f_{IKLJ}\omega^{BC}\omega^{AD}Z^K_BZ^L_CZ^J_D
-\frac{2}{3}f_{IKLJ}\gamma^{BC}\gamma^{AD}Z^K_BZ^L_CZ^J_D.
\end{eqnarray}
Now it is straightforward to calculate the bosonic potential:
\begin{eqnarray}\label{PotB}
-\frac{1}{2}\bar{F}^A_IF^I_A&=&\frac{1}{18}f_{IJKO}f^{O}{}_{LMN}
(-\omega^{AC}\omega^{BE}\omega^{DF}
+2\omega^{AC}\gamma^{BE}\gamma^{DF}\nonumber\\
&&+2\omega^{DF}\gamma^{AC}\gamma^{BE}-
4\omega^{BE}\gamma^{AC}\gamma^{DF})Z^I_AZ^J_BZ^K_CZ^L_DZ^M_EZ^N_F.
\end{eqnarray}
Note that $V=\frac{1}{2}\bar{F}^A_IF^I_A$ is positive definite,
though it is not manifestly $Sp(4)$ invariant due to the presence
of the gamma matrix. However, by taking advantage of the key
identity (\ref{Sp4ID}) and the constraint condition
$f_{(IJK)L}=0$, we are able to prove that (\ref{PotB}) is indeed
$Sp(4)$ invariant (see appendix \ref{PotB2}). The final expression
for the bosonic potential is
\begin{equation}\label{PotB3}
V=-\frac{1}{60}(2f_{IJK}{}^Of_{OLMN}-9f_{KLI}{}^Of_{ONMJ}
+2f_{IJL}{}^Of_{OKMN})Z^N_AZ^{AI}Z^J_BZ^{BK}Z^L_CZ^{CM}.
\end{equation}
In summary, the full Lagrangian in terms of the symplectic 3-algebra
is given by
\begin{eqnarray}\label{GeneN5Lagran}\nonumber
{\cal L}&=&\frac{1}{2}(-D_\mu\bar{Z}^A_ID^\mu
Z^I_A+i\bar{\psi}^A_I\gamma_\mu D^\mu\psi^I_A)\nonumber\\
&&-\frac{i}{2}\omega^{AB}\omega^{CD}f_{IJKL}(Z^I_AZ^K_B\psi^J_C\psi^L_D-
2Z^I_AZ^K_D\psi^J_C\psi^L_B)\nonumber\\
&&+\frac{1}{2}\epsilon^{\mu\nu\lambda}(f_{IJKL}A_\mu^{IJ}\partial_\nu
A_\lambda^{KL}+\frac{2}{3}f_{IJK}{}^Of_{OLMN}A_\mu^{IJ}A_\nu^{KL}A_\lambda^{MN})\\
&&+\frac{1}{60}(2f_{IJK}{}^Of_{OLMN}-9f_{KLI}{}^Of_{ONMJ}+2f_{IJL}{}^Of_{OKMN})Z^N_A
Z^{AI}Z^J_BZ^{BK}Z^L_CZ^{CM}.\nonumber
\end{eqnarray}
This Lagrangian is exactly the same as the ${\cal N}=5$ Lagrangian
derived by requiring that the supersymmetry transformations are
closed on-shell \cite{Chen2}. Using the reality condition
(\ref{HermiCondiOnF}), one can recast the potential term into the
following form:
\begin{equation}\label{positivity}
V=\frac{2}{15}(\Upsilon^L_{ABC})^*\Upsilon^L_{ABC},
\end{equation}
where
\begin{equation}
\Upsilon^L_{ABC}\equiv f_{IJK}{}^L(Z^I_AZ^J_BZ^K_C
+\frac{1}{4}\omega_{BC}Z^I_AZ^J_DZ^{DK}).
\end{equation}
Now the potential term is manifestly positive definite.


Let us consider the supersymmetry transformations. The $\CN=1$
supersymmetry transformation of the scalar field is
\begin{equation}\label{N1}
\delta_QZ^I_A= i\epsilon^\alpha\gamma_A{}^B\psi^I_{\alpha B}.
\end{equation}
On the other hand, the action (\ref{GeneN5Lagran}) is invariant
under the $Sp(4)$ global symmetry transformation
\begin{eqnarray}
\delta_RZ^I_A=\Sigma_A{}^BZ^I_B\quad, \quad
\delta_R\psi^I_A=\Sigma_A{}^B\psi^I_B.
\end{eqnarray}
Therefore one can consider the commutator of $\delta_R$ and
$\delta_Q$:
\begin{equation}\label{RQ}
[\delta_R,\delta_Q]Z^I_A=i\epsilon^\alpha(
\gamma_A{}^B\Sigma_B{}^C-\Sigma_A{}^B\gamma_B{}^C)\psi^I_{\alpha C}.
\end{equation}
So the $\CN=1$ supersymmetry does \emph{not} commute with the
$Sp(4)$ global symmetry. Since the matrix $\gamma_A{}^B$ contains
four independent real parameters, equation (\ref{RQ}) suggests that
there are other 4 independent $\CN=1$ supersymmetries. Therefore one
may promote the $\CN=1$ supersymmetry (\ref{N1}) to $\CN=5$:
\begin{equation}
\delta Z^I_A= i\epsilon_A{}^{B\alpha}\psi^I_{B\alpha},
\end{equation}
where the parameter
$\epsilon_A{}^{B\alpha}=\epsilon^{\alpha}_m\gamma^{m}_A{}^B$. One
may apply the same argument to the supersymmetry transformations of
the fermionic and gauge fields. In summary, we have the following
supersymmetry transformations:
\begin{eqnarray}\label{GeneSusyTransLaw}\nonumber
\delta Z^I_A&=&i\epsilon_A{}^{B\alpha}\psi^I_{B\alpha},\nonumber\\
\delta\psi^I_{A\alpha}&=&(\gamma^{\mu})_\alpha{}^{\beta}D_\mu
Z^I_B\epsilon^B{}_{A\beta}
+\frac{1}{3}f^I{}_{JKL}\omega^{BC}Z^J_BZ^K_CZ^L_D\epsilon^D{}_{A\alpha}
-\frac{2}{3}f^I{}_{JKL}\omega^{BD}Z^J_CZ^K_DZ^L_A\epsilon^C{}_{B\alpha},
\nonumber\\
\delta \tilde{A}_\mu{}^K{}_L &=&
i\epsilon^{AB\alpha}(\gamma_\mu)_\alpha{}^{\beta}\psi^J_{B\beta}Z^I_Af_{IJ}{}^K{}_L,
\end{eqnarray}
where the parameter $\epsilon^{AB}$ is antisymmetric in $AB$,
satisfying
\begin{eqnarray}\label{SusyPara5}\nonumber
&&\omega_{AB}\epsilon^{AB}=0 ,\nonumber\\
&&\epsilon^{*}_{AB}=\omega^{AC}\omega^{BD}\epsilon_{CD}.
\end{eqnarray}
The supersymetry transformations are precisely the ones proposed in
Ref. \cite{Chen2}. 
 To verify the mechanism for enhancing the $\CN=1$ to $\CN=5$, it is
best to check the closure of (\ref{GeneSusyTransLaw}).
Fortunately, the closure of (\ref{GeneSusyTransLaw}) has been
checked explicitly in Ref. \cite{Chen2}: they are indeed closed
on-shell, and the corresponding equations of motion can be derived
from the Lagrangian (\ref{GeneN5Lagran}). So the R-symmetry of the
theories is $Sp(4)$.

\section{The $\CN=4$ Theories and Symplectic Three-Algebras }\label{secN4}
\subsection{$\CN=4$ Theories by Starting from $\CN=5$
Theories}\label{secn4.1} In this section, we will construct the
${\cal N}$=4 theories by decomposing the ${\cal N}=5$
supermultiplets and the symplectic 3-algebra properly and
proposing a new superpotential term that preserving only $\CN=4$.
Let us first decompose the $\CN=5$ super-fields for matter fields
into $\CN=4$ super-fields:
\begin{equation}\label{SS4}
(\Phi^I_A)_{\CN=5}=\begin{pmatrix}\Phi^a_A \\ \Phi^\pa_\DA
\end{pmatrix}=\begin{pmatrix}Z^a_A \\ Z^\pa_\DA
\end{pmatrix}+i\begin{pmatrix}0 & \quad\sigma_A{}^\DA \\
\sigma^\dag_\DA{}^A & \quad 0
\end{pmatrix}\begin{pmatrix}\psi^\pa_A \\ \psi^a_\DA
\end{pmatrix}-\frac{i}{2}\theta^2\begin{pmatrix}F^a_A \\ F^\pa_\DA
\end{pmatrix}.
\end{equation}
The index $A$ of the LHS runs from 1 to 4, while $A$ and $\DA$ of
the RHS run from 1 to 2. (For the dotted and un-dotted
representation, see Appendix \ref{SO4}.) The indices $a$ and $\pa$
run from 1 to $2M$ and 1 to $2N$, respectively. The superfields
$\Phi^a_\DA$ and $\Phi^\pa_\DA$ are called untwisted and twisted
hyper-multiplets, repsectively, in the literature
\cite{Hosomichi:2008jb} (from the $\CN=4$ point of view).
%
%
The two antisymmetric matrices $\omega^{IJ}$ and $\omega^{AB}$ are
decomposed as
\begin{equation}
\omega^{IJ}=\begin{pmatrix} \omega^{ab} & 0 \\
0 & \omega^{\pa\pb}
\end{pmatrix}\quad {\rm and} \quad \omega^{AB}=\begin{pmatrix} \epsilon^{AB} & 0 \\
0 & \epsilon^{\dot{A}\dot{B}}
\end{pmatrix}
\end{equation}
respectively. Now the reality condition
$(\bar{\Phi}^A_I)_{\CN=5}=\omega^{AB}\omega_{IJ}\Phi^J_B$ becomes
\begin{equation}
\bar{\Phi}^A_a=\epsilon^{AB}\omega_{ab}\Phi^b_B \quad {\rm and}
\quad
\bar{\Phi}^\DA_\pa=\epsilon^{\DA\DB}\omega_{\pa\pb}\Phi^\pb_\DB.
\end{equation}
To be compatible with the decomposition of the $\CN=5$
hype-multiplets (\ref{SS4}), one may decompose the $\CN=5$
super-connections as
\begin{equation}\label{Sconn4}
\Gamma^{IJ}f_{IJ}{}^K{}_L=\begin{pmatrix}\Gamma^{ab}f_{ab}{}^c{}_d
+\Gamma^{\pa\pb}f_{\pa\pb}{}^c{}_d &\quad 0\\
0 &\quad \Gamma^{\pa\pb}f_{\pa\pb}{}^\pc{}_\pd+
\Gamma^{ab}f_{ab}{}^\pc{}_\pd
\end{pmatrix},
\end{equation}
where
\begin{eqnarray}
\Gamma^{ab}f_{ab}{}^c{}_d=(i\theta^\beta A_{\alpha\beta}^{ab}
+\theta^2\chi_{\alpha}^{ab})f_{ab}{}^c{}_d,
\end{eqnarray}
and the other 3 superfields of the RHS of (\ref{Sconn4}) have
similar expressions. In proposing (\ref{Sconn4}), we have
decomposed the set of 3-algebra generators $T_I$ into two sets of
generators $T_a$ and $T_\pa$, and decomposed the 3-bracket
(\ref{Symp3Bracket}) into 4 sets, with the structure constants
$f_{abc}{}^d, f_{ab\pc}{}^\pd, f_{\pa\pb c}{}^d$ and
$f_{\pa\pb\pc}{}^\pd$. We have also decomposed the parameter
superfield $\Gamma^{IJ}$ into two superfields $\Gamma^{ab}$ and
$\Gamma^{\pa\pb}$.

If we introduce a `spin up' spinor $\chi_{1\alpha}$ and a `spin
down' spinor $\chi_{2\alpha}$, i.e., \footnote{Here the index $\a$
is \emph{not} an index of a spacetime spinor. We hope this will not
cause any confusion.}
\begin{eqnarray}
\chi_{1\alpha}=\begin{pmatrix} 1 \\ 0
\end{pmatrix}=\d_{1\a}\;
\;\;{\rm and}\;\;\; \chi_{2\alpha}=\begin{pmatrix} 0 \\ 1
\end{pmatrix}=\d_{2\a},
\end{eqnarray}
then in component formalism, we now have
\begin{equation}\label{DCPSf5}
f_{IJKL}=f_{abcd}\d_{1\a}\d_{1\b}\d_{1\g}\d_{1\d}
+f_{ab\pc\pd}\d_{1\a}\d_{1\b}\d_{2\g}\d_{2\d}+f_{\pa\pb
cd}\d_{2\a}\d_{2\b}\d_{1\g}\d_{1\d}+f_{\pa\pb\pc\pd}\d_{2\a}\d_{2\b}\d_{2\g}\d_{2\d},
\end{equation}
(Here we assume that $(f_{abcd}-f_{ab\pc\pd})$ does not vanish
identically.) and
\e\label{n4conn}
\Gamma^{IJ}=\Gamma^{ab}\d_{1\a}\d_{1\b}+\Gamma^{\pa\pb}\d_{2\a}\d_{2\b}.
\ee
Substituting (\ref{DCPSf5}) and (\ref{n4conn}) into
$\Gamma^{IJ}f_{IJ}{}^K{}_L$ indeed gives (\ref{Sconn4}). With the
decomposition (\ref{DCPSf5}), the FI (\ref{FFI}) are decomposed
into 4 sets:
\e \label{FI4}&&f_{abe}{}^gf_{gfcd}+
f_{abf}{}^gf_{egcd}-f_{efd}{}^gf_{abcg}-f_{efc}{}^gf_{abdg}=0 ,\nonumber\\
&&f_{abe}{}^gf_{gf\pc\pd}+ f_{abf}{}^gf_{eg\pc\pd}-f_{ef\pd}{}^\pg
f_{ab\pc\pg}-f_{ef\pc}{}^\pg f_{ab\pd\pg}=0 ,\\
&&f_{\pa\pb e}{}^gf_{gf\pc\pd}+ f_{\pa\pb
f}{}^gf_{eg\pc\pd}-f_{ef\pd}{}^\pg f_{\pa\pb\pc\pg}-f_{ef\pc}{}^\pg
f_{\pa\pb\pd\pg}=0 ,\nonumber\\&&f_{\pa\pb \pe}{}^\pg
f_{\pg\pf\pc\pd}+ f_{\pa\pb \pf}{}^\pg
f_{\pe\pg\pc\pd}-f_{\pe\pf\pd}{}^\pg
f_{\pa\pb\pc\pg}-f_{\pe\pf\pc}{}^\pg f_{\pa\pb\pd\pg}=0
.\nonumber\ee
In accordance with Eq. (\ref{SymmeOfF}), these structure constants
enjoy the symmetry properties
\begin{eqnarray}\label{symfs}
&&f_{abcd}=f_{bacd}=f_{badc}=f_{cdab},\nonumber\\
&&f_{ab\pc\pd}=f_{ba\pc\pd}=f_{ba\pd\pc}=f_{\pc\pd ab},\\
&&f_{\pa\pb\pc\pd}=f_{\pb\pa\pc\pd}=f_{\pb\pa\pd\pc}=f_{\pc\pd\pa\pb}.\nonumber
\end{eqnarray}
The reality condition (\ref{HermiCondiOnF}) is decomposed into
\begin{equation}f^{*a}{}_b{}^c{}_d=f^{b}{}_a{}^d{}_c,
\quad f^{*\pa}{}_\pb{}^c{}_d=f^{\pb}{}_\pa{}^d{}_c,\quad
f^{*\pa}{}_\pb{}^\pc{}_\pd=f^{\pb}{}_\pa{}^\pd{}_\pc.
\end{equation}
Under the condition that $(f_{abcd}-f_{ab\pc\pd})$ does not vanish
identically, decomposing the constraint condition $f_{(IJK)L}=0$
results in $f_{(abc)d}=0$, $f_{(\pa\pb\pc)\pd}=0$ and
$f_{ab\pc\pd}=0$. However, the condition $f_{ab\pc\pd}=0$ turns out
to be too restrictive to allow any interaction between the primed
fields and the un-primed fields. So we have to give up the
constraint $f_{ab\pc\pd}=0$. Namely, we have to give up the
constraint condition $f_{(IJK)L}=0$ as we decompose $f_{IJKL}$ by
Eq. (\ref{DCPSf5}). Later we will see, to construct an interesting
$\CN=4$ quiver gauge theory, we need only to impose constraints on
$f_{abcd}$ and $f_{\pa\pb\pc\pd}$:
\begin{equation}\label{Constr3}
f_{(abc)d}=0\quad {\rm and} \quad f_{(\pa\pb\pc)\pd}=0,
\end{equation}
while $f_{ab\pc\pd}$ are un-constrained.

With these decompositions, the Lagrangian for the kinetic terms of
the matter fields (\ref{Lkin}) becomes \e\label{Lkin4} {\cal
L}_{{\rm kin}}&=&\frac{1}{2}(-D_\mu\bar{Z}^A_aD^\mu
Z^a_A+i\bp^\DA_a\g^\mu
D_\mu\p^a_\DA-2i\s^\dag_\DB{}^A\bp^\DB_a\tilde{\chi}^a{}_bZ^b_A+\bar{F}^A_aF^a_A)\nonumber\\
&&+\frac{1}{2}(-D_\mu\bar{Z}^\DA_\pa D^\mu
Z^\pa_\DA+i\bp^A_\pa\g^\mu
D_\mu\p^\pa_A-2i\s_B{}^\DA\bp^B_\pa\tilde{\chi}^\pa{}_\pb
Z^\pb_\DA+\bar{F}^\DA_\pa F^\pa_\DA),
\ee
where
\e D_\mu Z^A_d &=&
\partial_\mu Z^A_d -\tilde A_\mu{}^c{}_dZ^A_c ,\nonumber\\
\tilde A_\mu{}^c{}_d&=&A^{ab}_\mu f_{ab}{}^c{}_d+A^{\pa\pb}_\mu
f_{\pa\pb}{}^c{}_d ,\nonumber\\ \tilde\chi{}^\pa{}_\pb&=&\chi^{cd}
f_{cd}{}^\pa{}_\pb+\chi^{\pc\pd}f_{\pc\pd}{}^\pa{}_\pb, \ee
and similar definitions for $\tilde A_\mu{}^\pc{}_\pd$ and
$\tilde\chi{}^a{}_b$; and the Chern-Simons term (\ref{LCS}) becomes
\e\label{LCS4} \CL_{{\rm
CS}}&=&\frac{1}{2}\epsilon^{\mu\nu\lambda}(f_{abcd}A_\mu^{ab}\partial_\nu
A_\lambda^{cd}+\frac{2}{3}f_{abc}{}^gf_{gdef}A_\mu^{ab}A_\nu^{cd}A_\lambda^{ef})\nonumber\\
&&+\frac{1}{2}\epsilon^{\mu\nu\lambda}(f_{\pa\pb\pc\pd}A_\mu^{\pa\pb}\partial_\nu
A_\lambda^{\pc\pd}+\frac{2}{3}f_{\pa\pb\pc}{}^\pg
f_{\pg\pd\pe\pf}A_\mu^{\pa\pb}A_\nu^{\pc\pd}A_\lambda^{\pe\pf})\\
&&+\epsilon^{\mu\nu\lambda}(f_{ab\pc\pd}A_\mu^{ab}\partial_\nu
A_\lambda^{\pc\pd}+f_{abc}{}^g
f_{gd\pe\pf}A_\mu^{ab}A_\nu^{cd}A_\lambda^{\pe\pf}+f_{ab\pc}{}^\pg
f_{\pg\pd\pe\pf}A_\mu^{ab}A_\nu^{\pc\pd}A_\lambda^{\pe\pf})\nonumber\\
&&+\frac{i}{2}(f_{abcd}\chi^{ab}\chi^{cd}+2f_{ab\pc\pd}\chi^{ab}\chi^{\pc\pd}
+f_{\pa\pb\pc\pd}\chi^{\pa\pb}\chi^{\pc\pd})\nonumber. \ee
The equations of motion for the auxiliary field $\chi$ (\ref{Chi})
is decomposed into two sets
\e\label{Chi4}&&\chi^{ab}=-\s^\dag_\DA{}^B\p^{\DA (a}Z^{b)}_B
,\nonumber\\ &&\chi^{\pa\pb}=-\s_A{}^\DB\p^{A (\pa}Z^{\pb)}_\DB .
\ee
Plugging $(\ref{Chi4})$ into (\ref{Lkin4}) and (\ref{LCS4}) gives
three Yukawa terms
\e\label{YKW41}
&&-\frac{i}{2}(f_{acbd}\s^{A\DC}\s^{B\DD}Z^a_AZ^b_B\p^c_\DC\p^d_\DD
+f_{\pa\pc\pb\pd}\s^{\dag\DA C}\s^{\dag\DB D}Z^\pa_\DA
Z^\pb_\DB\p^\pc_C\p^\pd_D
\nonumber\\&&\quad\quad+2f_{ab\pc\pd}\s^{A\DB}\s^{\dag\DC
D}Z^a_AZ^\pc_\DC\p^b_\DB\p^\pd_D ). \ee
Alternatively, we can also obtain (\ref{YKW41}) by directly
decomposing the $\CN=5$ Yukawa term (\ref{YKW1}). It can be seen
that the last term of (\ref{YKW41}) is a mixed term, in which the
primed fields couple the un-primed fields through $f_{ab\pc\pd}$. So
we cannot obtain a non-trivial $\CN=4$ superpotential by decomposing
the $\CN=5$ superpotential (\ref{SuperPot}), because the $\CN=5$
superpotential (\ref{SuperPot}) is desired only if $f_{(IJK)L}=0$,
which implies that $f_{ab\pc\pd}=0$ as we decompose $f_{IJKL}$ by
Eq. (\ref{DCPSf5}) under the condition that
$(f_{abcd}-f_{ab\pc\pd})$ does not vanish identically. So we have to
propose a new superpotential for the $\CN=4$ theory, allowing
$f_{ab\pc\pd}\neq 0$. However, unlike the last term of
(\ref{YKW41}), the first two terms of (\ref{YKW41}) are un-mixed
terms. This inspires us to decompose the first term of the $\CN=5$
superpotential (\ref{SuperPot0}) with $f_{\pa\pc bd}$ and
$f_{ac\pb\pd}$ deleted from $f_{IKJL}$ (hence we denote the
`modified' structure constants as $f^\prime_{IKJL}$):
\e \label{superPot41}W_1(\Phi)&=&
\frac{1}{12}(f^\prime_{IKJL}\omega^{AB}\omega^{CD}\Phi^{I}_A\Phi^{J}_B\Phi^{K}_C\Phi^{L}_D)_{\CN=5}\nonumber\\
&=&\frac{1}{12}(f_{acbd}\ep^{AB}\ep^{CD}\Phi^{a}_A\Phi^{b}_B\Phi^{c}_C\Phi^{d}_D+
f_{\pa\pc\pb\pd}\ep^{\DA\DB}\ep^{\DC\DD}\Phi^{\pa}_\DA\Phi^{\pb}_\DB\Phi^{\pc}_\DC\Phi^{\pd}_\DD).\ee
where
\begin{eqnarray}
f^\prime_{IJKL}=f_{abcd}\d_{1\a}\d_{1\b}\d_{1\g}\d_{1\d}
+f_{\pa\pb\pc\pd}\d_{2\a}\d_{2\b}\d_{2\g}\d_{2\d}.
\end{eqnarray}
Of course, the `modified' structure constants $f^\prime_{IKJL}$
still satisfy the constraint condition $f^\prime_{(IKJ)L}$=0, which
is equivalent to Eq. (\ref{Constr3}): $f_{(acb)d}=0$ and
$f_{(\pa\pc\pb)\pd}=0$. We will prove that the first two terms of
(\ref{YKW41}) combining the Yukawa terms arising from the
superpotential $W_1$ (see (\ref{superPot41})) are $SU(2)\times
SU(2)$ invariant. Carrying out the Berezin integral $\frac{i}{2}\int
d\theta^2W_1(\Phi)$ gives
\e\label{superPot41}
\CL_{W_1}&=&-\frac{i}{6}(f_{acbd}\ep^{AB}\ep^{\DC\DD}Z^a_AZ^b_B\p^c_\DC\p^d_\DD
+f_{\pa\pc\pb\pd}\ep^{\DA\DB}\ep^{CD}Z^\pa_\DA
Z^\pb_\DB\p^\pc_C\p^\pd_D)\nonumber\\
&&-\frac{i}{6}[(f_{abcd}-f_{adcb})\s^{A\DC}\s^{B\DD}Z^a_AZ^b_B\p^c_\DC\p^d_\DD
+(f_{\pa\pb\pc\pd}-f_{\pa\pd\pc\pb})\s^{\dag \DA C}\s^{\dag\DB
D}Z^\pa_\DA
Z^\pb_\DB\p^\pc_C\p^\pd_D]\nonumber\\
&&-\frac{1}{3}(f_{abcd}Z^b_BZ^{Bc}Z^{Ad}F^a_A+f_{\pa\pb\pc\pd}Z^\pb_\DB
Z^{\DB\pc}Z^{\DA\pd}F^\pa_\DA).\ee
Let us now combine the first term of (\ref{YKW41}) and the first
term of the second line of (\ref{superPot41}):
\e\label{YKW42}
&&-\frac{i}{6}[3f_{acbd}+(f_{abcd}-f_{adcb})]\s^{A\DC}\s^{B\DD}Z^a_AZ^b_B\p^c_\DC\p^d_\DD\nonumber\\
&=&
-\frac{i}{6}(f_{acbd}-f_{bcad})(\s^{A\DC}\s^{B\DD}-\s^{B\DC}\s^{A\DD})Z^a_AZ^b_B\p^c_\DC\p^d_\DD\nonumber\\
&=&-\frac{i}{3}f_{acbd}\ep^{AB}\ep^{\DC\DD}Z^a_AZ^b_B\p^c_\DC\p^d_\DD.
\ee
In the second line we have used $f_{(abc)d}=0$. In the third line we
have used the $SU(2)\times SU(2)$ identity (\ref{SU2ID}). It can be
seen that the final expression of (\ref{YKW42}) is indeed
$SU(2)\times SU(2)$ invariant. Similarly, one can combine the second
term of (\ref{YKW41}) and the second term of the second line of
(\ref{superPot41}) to form an $SU(2)\times SU(2)$ invariant
expression:
\e\label{YKW43}
-\frac{i}{3}f_{\pa\pc\pb\pd}\ep^{\DA\DB}\ep^{CD}Z^\pa_\DA
Z^\pb_\DB\p^\pc_C\p^\pd_D, \ee
where we have used the reality condition (\ref{RC4}). Now only the
last term of (\ref{YKW41}), i.e. the mixed term, is not $SU(2)\times
SU(2)$ invariant. Its structure suggests that if a Yukawa term of
the form
\begin{equation}\label{YKW44}
if_{ab\pc\pd}\s^{D\DB}\s^{\dag\DC A}Z^a_AZ^\pc_\DC\p^b_\DB\p^\pd_D
\end{equation}
arises from a to-be-determined superpotential, then they will add up
to be $SU(2)\times SU(2)$ invariant by the reality condition
(\ref{RC4}) and the identity (\ref{SU2ID}). It is therefore natural
to try
\begin{equation}
W_2(\Phi)=\alpha f_{ab\pc\pd}\s^{B\DD}\s^{\dag\DC
A}\Phi^a_A\Phi^b_B\Phi^\pc_\DC\Phi^\pd_\DD ,
\end{equation}
where $\alpha$ is a constant, to be determined later. The
corresponding Lagrangian is
\e \label{YKW45}\CL_{W_2}&=&i\alpha
f_{ab\pc\pd}(\ep^{AC}\ep^{BD}Z^a_AZ^b_B\p^\pc_C\p^\pd_D+\ep^{\DA\DC}\ep^{\DB\DD}\p^a_\DA
\p^b_\DB Z^\pc_\DC Z^\pd_\DD+2\ep^{AC}\ep^{\DB\DD}Z^a_A
Z^\pd_\DD\p^b_\DB\p^\pc_C)\nonumber\\
&&+2i\alpha f_{ab\pc\pd}\s^{D\DB}\s^{\dag\DC A}Z^a_A
Z^\pc_\DC\p^b_\DB\p^\pd_D\nonumber\\&&-2\alpha
f_{ab\pc\pd}\s^{B\DD}\s^{\dag\DC A}Z^b_BZ^\pc_\DC Z^\pd_\DD
F^a_A-2\alpha f_{ab\pc\pd}\s^{B\DD}\s^{\dag\DC A}Z^a_AZ^b_B
Z^\pd_\DD F^\pc_\DC.\ee
Note that the first line is $SU(2)\times SU(2)$ invariant by itself.
Comparing the second line with (\ref{YKW44}) gives
$\alpha=\frac{1}{2}$. Combining the last term of (\ref{YKW41}) and
the second line of (\ref{YKW45}), we obtain
\begin{equation}\label{YKW46}
if_{ab\pc\pd}\ep^{AD}\ep^{\DB\DC}Z^a_A Z^\pc_\DC\p^b_\DB\p^\pd_D,
\end{equation}
which is the desired result. Now all Yukawa terms are invariant
under the $SU(2)\times SU(2)$ global symmetry transformation. Put
all Yukawa terms (the first line of (\ref{superPot41}),
(\ref{YKW42}), (\ref{YKW43}), (\ref{YKW46}) and the first line of
(\ref{YKW45})) together:
\e \label{YKWT4}\CL_{\rm Y}
&=&-\frac{i}{2}(f_{acbd}Z^a_AZ^{Ab}\p^c_\DB\p^{\DB d}+
f_{\pa\pc\pb\pd}Z^\pa_\DA Z^{\DA\pb}\p^\pc_B\p^{B\pd
})\nonumber\\&&+\frac{i}{2} f_{ab\pc\pd}(Z^a_AZ^b_B\p^{A\pc}\p^{
B\pd}+Z^{\pc}_\DA Z^{\pd}_\DB\p^{\DA a}\p^{\DB b}+4Z^a_A
Z^{\DB\pd}\p^b_\DB\p^{A\pc}).\nonumber\ee

To calculate the bosonic potential, we first integrate out the
auxiliary fields $F^a_A$ and  $F^\pa_\DA$ from (\ref{Lkin4}),
(\ref{superPot41}) and (\ref{YKW45}):
\begin{eqnarray}
&&\bar{F}^A_a=\frac{1}{3}f_{abcd}Z^b_BZ^{Bc}Z^{Ad}+f_{ab\pc\pd}\s^{B\DD}\s^{\dag\DC
A}Z^b_BZ^\pc_\DC Z^\pd_\DD\equiv W^A_{1a} + W^A_{2a} ,\nonumber\\
&&\bar{F}_\pa^\DA=\frac{1}{3}f_{\pa\pb\pc\pd}Z^\pb_\DB
Z^{\DB\pc}Z^{\DA\pd}+f_{\pa\pb cd}\s^{\dag\DB D}\s^{C\DA }Z^\pb_\DB
Z^c_CZ^d_D \equiv W^\DA_{1\pa} + W^\DA_{2\pa}.
\end{eqnarray}
The bosonic potential is
\e\label{Pot40} -V=-\frac{1}{2}(\bar{F}^A_aF^a_A+\bar{F}^\DA_\pa
F^\pa_\DA), \ee
which is not manifestly $SU(2)\times SU(2)$ invariant due to the
presence of the sigma matrices. However, by using the fundamental
identities (\ref{FI4}) and a method first introduced in GW theory
\cite{GaWi} (see also \cite{HosomichiJD}), we are able to re-write
(\ref{Pot40}) so that it has a manifest $SU(2)\times SU(2)$ global
symmetry. For example, let us consider
\e\label{Pot41} -W^A_{1a}W_{2A}^{a}&=&
-\frac{1}{3}f_{abcd}f^a{}_{e\pc\pd}\s^{A\DC} \s^{C\DD}
Z^b_BZ^{Bc}Z^{d}_AZ^e_CZ^\pc_\DC Z^\pd_\DD\nonumber\\&=&
-\frac{1}{3}\{f_{cda(b}f^a{}_{e)\pc\pd}+f_{cda[b}f^a{}_{e]\pc\pd}\}\s^{A\DC}
\s^{C\DD} Z^b_BZ^{Bc}Z^{d}_AZ^e_CZ^\pc_\DC
Z^\pd_\DD\nonumber\\&\equiv& S+A . \ee
The antisymmetric part can be written as
\e A=\frac{1}{6}f_{cdab}f^a{}_{e\pc\pd}\s^{A\DC} \s^{C\DD}
Z^{Bb}Z^{e}_BZ^{c}_CZ^{d}_AZ^\pc_\DC Z^\pd_\DD . \ee
Applying the constraint condition $f_{(cdb)a}=0$ to the above
potential term, we obtain
\e A=-\frac{1}{3}f_{cdae}f^a{}_{b\pc\pd}\s^{A\DC} \s^{C\DD}
Z^b_BZ^{Bc}Z^{d}_AZ^e_CZ^\pc_\DC Z^\pd_\DD .\ee
Combining this with $-W^A_{1a}W_{2A}^{a}$ (the first line of
(\ref{Pot41})) gives
\begin{eqnarray}
-W^A_{1a}W_{2A}^{a}+A=2S.
\end{eqnarray}
Solving for $-W^A_{1a}W_{2A}^{a}$, we obtain
\begin{eqnarray}\label{pot42}
-W^A_{1a}W_{2A}^{a}=-\frac{1}{2}f_{cda(b}f^a{}_{e)\pc\pd}\s^{A\DC}
\s^{C\DD} Z^b_BZ^{Bc}Z^{d}_AZ^e_CZ^\pc_\DC Z^\pd_\DD.
\end{eqnarray}
Let us now consider another term of (\ref{Pot40}):
\e-\frac{1}{2}W^\DA_{2\pa}W_{2\DA}^{\pa}&=&\frac{1}{2}f_{cd
\pb\pa}f^\pa{}_{\pe fg}\s^{D\DB} \s^{A\dot F} Z^\pb_\DB
Z^{\pe}_{\dot F} Z^{c}_C Z^d_D Z^{Cf} Z^{g}_A\\
&=&\frac{1}{2}(f_{cd\pa(\pb}f^\pa{}_{\pe)fg}+f_{cd\pa[\pb}f^\pa{}_{\pe]fg})\s^{D\DB}
\s^{A\dot F} Z^\pb_\DB Z^{\pe}_{\dot F} Z^{c}_C Z^d_D Z^{Cf} Z^{g}_A
.\nonumber\ee
Combining this equation with (\ref{pot42}), the symmetric part
cancels (\ref{pot42}) by the second equation of the fundamental
identities (\ref{FI4}), while the antisymmetric part is $SU(2)\times
SU(2)$ invariant by the identity (\ref{SU2ID}). The final result is
\begin{equation}
-W^A_{1a}W_{2A}^{a}-\frac{1}{2}W^\DA_{2\pa}W_{2\DA}^{\pa}=-\frac{1}{4}f_{ab
\pc\pg}f^\pg{}_{\pd ef}Z^{\DA\pc}Z^\pd_\DA Z^b_D Z^{Df}Z^a_C Z^{Ce}.
\end{equation}
One can apply the same method to the other terms of (\ref{Pot40}).
The final expression for the $\CN=4$ bosonic potential is
\e-V&=&+\frac{1}{12}(f_{abcg}f^g{}_{def}Z^{Aa}Z^b_BZ^{B(c}Z^{d)}_CZ^{Ce}Z^f_A
+f_{\pa\pb\pc\pg}f^\pg{}_{\pd\pe\pf}Z^{\DA\pa}Z^\pb_\DB
Z^{\DB(\pc}Z^{\pd)}_\DC Z^{\DC\pe}Z^\pf_\DA)\nonumber\\
&&-\frac{1}{4}(f_{ab \pc\pg}f^\pg{}_{\pd ef}Z^{\DA\pc}Z^\pd_\DA
Z^b_D Z^{Df}Z^a_C Z^{Ce}+f_{\pa\pb
cg}f^g{}_{d\pe\pf}Z^{Ac}Z^d_AZ^\pb_\DD Z^{\DD\pf}Z^\pa_\DC
Z^{\DC\pe})\nonumber\\ \ee
In summary, the full $\CN=4$ Lagrangian is given by
\e\label{LN4}\CL&=&\frac{1}{2}(-D_\mu\bar{Z}^A_aD^\mu
Z^a_A-D_\mu\bar{Z}^\DA_\pa D^\mu Z^\pa_\DA+i\bp^\DA_a\g^\mu
D_\mu\p^a_\DA+i\bp^A_\pa\g^\mu
D_\mu\p^\pa_A)\nonumber\\&&-\frac{i}{2}(f_{acbd}Z^a_AZ^{Ab}\p^c_\DB\p^{\DB
d}+ f_{\pa\pc\pb\pd}Z^\pa_\DA Z^{\DA\pb}\p^\pc_B\p^{B\pd
})\nonumber\\&&+\frac{i}{2} f_{ab\pc\pd}(Z^a_AZ^b_B\p^{A\pc}\p^{
B\pd}+Z^{\pc}_\DA Z^{\pd}_\DB\p^{\DA a}\p^{\DB b}+4Z^a_A
Z^{\DB\pd}\p^b_\DB\p^{A\pc})\nonumber\\&&
+\frac{1}{2}\epsilon^{\mu\nu\lambda}(f_{abcd}A_\mu^{ab}\partial_\nu
A_\lambda^{cd}+\frac{2}{3}f_{abc}{}^gf_{gdef}A_\mu^{ab}A_\nu^{cd}A_\lambda^{ef})\nonumber\\
&&+\frac{1}{2}\epsilon^{\mu\nu\lambda}(f_{\pa\pb\pc\pd}A_\mu^{\pa\pb}\partial_\nu
A_\lambda^{\pc\pd}+\frac{2}{3}f_{\pa\pb\pc}{}^\pg
f_{\pg\pd\pe\pf}A_\mu^{\pa\pb}A_\nu^{\pc\pd}A_\lambda^{\pe\pf})\nonumber\\
&&+\epsilon^{\mu\nu\lambda}(f_{ab\pc\pd}A_\mu^{ab}\partial_\nu
A_\lambda^{\pc\pd}+f_{abc}{}^g
f_{gd\pe\pf}A_\mu^{ab}A_\nu^{cd}A_\lambda^{\pe\pf}+f_{ab\pc}{}^\pg
f_{\pg\pd\pe\pf}A_\mu^{ab}A_\nu^{\pc\pd}A_\lambda^{\pe\pf})\nonumber\\
&&+\frac{1}{12}(f_{abcg}f^g{}_{def}Z^{Aa}Z^b_BZ^{B(c}Z^{d)}_CZ^{Ce}Z^f_A
+f_{\pa\pb\pc\pg}f^\pg{}_{\pd\pe\pf}Z^{\DA\pa}Z^\pb_\DB
Z^{\DB(\pc}Z^{\pd)}_\DC Z^{\DC\pe}Z^\pf_\DA)\nonumber\\
&&-\frac{1}{4}(f_{ab \pc\pg}f^\pg{}_{\pd ef}Z^{\DA\pc}Z^\pd_\DA
Z^b_D Z^{Df}Z^a_C Z^{Ce}+f_{\pa\pb
cg}f^g{}_{d\pe\pf}Z^{Ac}Z^d_AZ^\pb_\DD Z^{\DD\pf}Z^\pa_\DC
Z^{\DC\pe}).\nonumber\\ \ee
Using the same argument given in Sec. \ref{N1to5}, we may promote
the $\CN=1$ supersymmetry transformations to $\CN=4$:
\e \label{SUSY4}&&\delta Z^a_A=i\ep_A{}^\DA\p^a_\DA,\nonumber\\
&&\delta Z^\pa_\DA=i\ep^\dag_\DA{}^A\p^\pa_A,\nonumber\\
&&\delta\p^\pa_A=-\g^\mu D_\mu
Z^\pa_\DB\ep_A{}^\DB-\frac{1}{3}f^\pa{}_{\pb\pc\pd}Z^\pb_\DB
Z^{\DB\pc}Z^\pd_\DC\ep_A{}^\DC+f^\pa{}_{\pb cd}Z^\pb_\DA
Z^{Bc}Z^d_A\ep_B{}^\DA, \nonumber\\
&&\delta\p^a_\DA=-\g^\mu D_\mu
Z^a_B\ep^\dag_\DA{}^B-\frac{1}{3}f^a{}_{bcd}Z^b_B
Z^{Bc}Z^d_C\ep^\dag_\DA{}^C+f^a{}_{b \pc\pd}Z^b_A
Z^{\DB\pc}Z^\pd_\DA\ep^\dag_\DB{}^A,\nonumber\\
&&\delta\tilde A_\mu{}^c{}_d=i\ep^{A\DB}\g_\mu\p^b_\DB
Z^a_Af_{ab}{}^c{}_d+i\ep^{\dag\DA B}\g_\mu\p^\pb_BZ^\pa_\DA
f_{\pa\pb}{}^c{}_d,\nonumber\\
&&\delta\tilde A_\mu{}^\pc{}_\pd=i\ep^{A\DB}\g_\mu\p^b_\DB
Z^a_Af_{ab}{}^\pc{}_\pd+i\ep^{\dag\DA B}\g_\mu\p^\pb_BZ^\pa_\DA
f_{\pa\pb}{}^\pc{}_\pd ,\ee
where the parameter satisfies the reality condition
\begin{equation}\label{n4para}
\ep^{\dag}{}_{\dot{A}}{}^{B}=
-\epsilon^{BC}\epsilon_{\dot{A}\dot{B}}\ep_{C}{}^{\dot{B}}.
\end{equation}
It is still necessary to verify the closure of the $\CN=4$
superalgebra; this will be done in the next subsection. The ordinary
Lie algebra counterparts of the Lagrangian (\ref{LN4}) and the
supersymmetry transformaitons (\ref{SUSY4}) are first constructed in
Ref. \cite{HosomichiJD}. If $f_{ab\pc\pd}=f_{abcd}$, then
$f_{ab\pc\pd}$ also satisfy the constraint equation, i.e.
$f_{(ab\pc)\pd}=0$. In this special case, the $\CN=4$ supersymmetry
will be enhanced to $\CN=5$. 
Therefore without the twisted hypermultiplet, it is impossible to enhance the $\CN=4$
supersymmetry to $\CN=5$; as a result, the $\CN=4$ supersymmetry cannot be promoted to $\CN=6,8$. Indeed, in Ref. \cite{HosomichiJD}, it was demonstrated that  the $\CN=4$ theory with an $SU(2)\times SU(2)$ gauge group is equivalent to the $
\CN=8$ BLG theory \emph{after} adding the twisted hypermultiplet.

In a forthcoming paper \cite{ChenWunew}, we will convert the $\CN=4$ theories (based on the 3-algebras) into general $\CN=4$ theories in terms of ordinary Lie (2-)algebras, using superalgebras to realize the 3-algebras. The method will be generalized to construct $\CN=4$ quiver gauge theories \cite{ChenWunew}. There are a special class of $\CN=4$ theories, with a circular quiver gauge diagram \cite{HosomichiJD, ChenWunew}:
\begin{equation}\label{linearN}
\cdots-U(N_{i-1})-U(N_i)-U(N_{i+1})-\cdots.
\end{equation}
(The above diagram is only a part of the circular quiver gauge diagram.) This class of $\CN=4$ theories have been conjectured to be the gauge descriptions of multi
M2-branes in orbifold $(\textbf{C}^2/\textbf{Z}_p\times \textbf{C}^2/\textbf{Z}_q)/\textbf{Z}_k$, where $p$ ($q$) is the number of the un-twisted (twisted) hypermultiplets, and $k$ the Chern-Simons level \cite{Imamura}.
Their gravity duals have been investigated in Ref. \cite{Imamura}. To our knowledge, most of the gravity duals of the $\CN=4$ quiver gauge theories are not found yet.
We would like to construct their gravity duals in the future. 

If one sets the twisted hypermultiplet to be zero, i.e.,
$\Phi^\pa_\DA=0$, then (\ref{LN4}) and (\ref{SUSY4}) become the
Lagrangian and the supersymmetry law of the GW theory \cite{GaWi},
respectively, in the 3-algebra approach:
\e\label{LN4GW}\CL&=&\frac{1}{2}(-D_\mu\bar{Z}^A_aD^\mu
Z^a_A+i\bp^\DA_a\g^\mu
D_\mu\p^a_\DA)-\frac{i}{2}f_{acbd}Z^a_AZ^{Ab}\p^c_\DB\p^{\DB
d}\nonumber\\&&+\frac{1}{2}\epsilon^{\mu\nu\lambda}(f_{abcd}A_\mu^{ab}\partial_\nu
A_\lambda^{cd}+\frac{2}{3}f_{abc}{}^gf_{gdef}A_\mu^{ab}A_\nu^{cd}A_\lambda^{ef})\nonumber\\
&&+\frac{1}{12}f_{abcg}f^g{}_{def}Z^{Aa}Z^b_BZ^{B(c}Z^{d)}_CZ^{Ce}Z^f_A,
\ee
and
\e \label{SUSY4GW}&&\delta Z^a_A=i\ep_A{}^\DA\p^a_\DA,\nonumber\\
&&\delta\p^a_\DA=-\g^\mu D_\mu
Z^a_B\ep^\dag_\DA{}^B-\frac{1}{3}f^a{}_{bcd}Z^b_B
Z^{Bc}Z^d_C\ep^\dag_\DA{}^C,\nonumber\\
&&\delta\tilde A_\mu{}^c{}_d=i\ep^{A\DB}\g_\mu\p^b_\DB
Z^a_Af_{ab}{}^c{}_d.\nonumber\\
\ee
\subsection{Closure of the \CN=4
Algebra}\label{CloseN4} 
The closure of the algebra of the GW theory was checked in
\cite{GaWi}. To our knowledge, there is no explicit check in the
literature for the closure of the $\CN=4$ algebra after adding the
twisted hypermultiplets into the GW theory. Here we present such a
check by starting with the supersymmetry transformation of the
scalar fields:
\e \label{STS4} [\delta_1, \delta_2]Z_A^a&=&v^\mu D_\mu
Z^a_A+\frac{1}{3}f^a{}_{bcd}Z^b_BZ^c_CZ^d_D\ep_{AE}\ep^{BC}u^{ED}\nonumber\\
&&+if^a{}_{b\pc\pd}Z^{Bb}Z^{\DB\pc}Z^{\DA\pd}(\ep_{2A\DA}\ep^\dag_{1\DB
B}-\ep_{1A\DA}\ep^\dag_{2\DB B}), \ee
where
\begin{equation}
v^\mu\equiv i\ep^\dag_{1\DA}{}^B\g^\mu\ep_{2B}{}^\DA,\quad\quad
u^{ED}\equiv
i(\ep_1^{E\DA}\ep^{\dag}_{2\DA}{}^D-\ep_2^{E\DA}\ep^{\dag}_{1\DA}{}^D).
\end{equation}
By using the identity
$\ep_{AE}\ep^{BC}=-(\delta_A{}^B\delta_E{}^C-\delta_E{}^B\delta_A{}^C)$,
the second term of the RHS of (\ref{STS4}) can be written as
\begin{equation}
-\frac{1}{3}f^a{}_{bcd}Z^b_AZ^c_CZ^d_Du^{CD}
+\frac{1}{3}f^a{}_{bcd}Z^b_BZ^c_AZ^d_Du^{BD}.
\end{equation}
The second term is equal to the first term minus the second term by
the constraint condition $f^a{}_{(bcd)}=0$:
\begin{equation}
\frac{1}{3}f^a{}_{bcd}Z^b_BZ^c_AZ^d_Du^{BD}=
-\frac{1}{3}f^a{}_{bcd}Z^b_AZ^c_CZ^d_Du^{CD}
-\frac{1}{3}f^a{}_{bcd}Z^b_BZ^c_AZ^d_Du^{BD}.
\end{equation}
Therefore the second term of the RHS of (\ref{STS4}) is equal to
\begin{equation}
-\frac{1}{2}f^a{}_{bcd}Z^c_CZ^d_Du^{CD}Z^b_A.
\end{equation}
By using the fourth equation of (\ref{SO4ID}), the second line of
the RHS of (\ref{STS4}) becomes
\begin{equation}
-\frac{1}{2}f^a{}_{b\pc\pd}Z^\pc_\DA Z^\pd_\DB u^{\DA\DB}Z^b_A,
\end{equation}
where
\begin{equation}
 u^{\DA\DB}\equiv i(\ep_1^{\dag\DA C}\ep_{2C}{}^\DB-\ep_2^{\dag\DA
 C}\ep_{1C}{}^\DB).
\end{equation}
In summary, we have
\e  [\delta_1, \delta_2]Z_A^a&=&v^\mu D_\mu
Z^a_A+\tilde\Lambda^a{}_bZ^b_A. \ee
While the first is the familiar covariant derivative, the second
term is a gauge transformation by a parameter
\begin{equation}
\tilde\Lambda^a{}_b\equiv-\frac{1}{2}f^a{}_{bcd}Z^c_CZ^d_Du^{CD}
-\frac{1}{2}f^a{}_{b\pc\pd}Z^\pc_\DA Z^\pd_\DB u^{\DA\DB}.
\end{equation}
Similarly, we have
\e  [\delta_1, \delta_2]Z_\DA^\pa&=&v^\mu D_\mu
Z^\pa_\DA+\tilde\Lambda^\pa{}_\pb Z^\pb_\DA, \ee
where the parameter $\tilde\Lambda^\pa{}_\pb $ is defined as
\begin{equation}
\tilde\Lambda^\pa{}_\pb\equiv-\frac{1}{2}f^\pa{}_{\pb\pc\pd}Z^\pc_\DC
Z^\pd_\DD u^{\DC\DD}-\frac{1}{2}f^\pa{}_{\pb cd}Z^c_A Z^d_B u^{AB}.
\end{equation}
Let us now examine the supersymmetry transformation of the gauge
fields:
\begin{eqnarray}
[\delta_1, \delta_2]\tilde{A}_\mu{}^a{}_b&=& v^\nu\tilde{F}_{\nu\mu}{}^a{}_b
-D_\mu\tilde\Lambda^{a}{}_b \nonumber\\
&&+v^\nu\{\tilde{F}_{\mu\nu}{}^a{}_b-\varepsilon_{\mu\nu\lambda}
[(Z^c_AD^\lambda\bar Z^{Ad}-\frac{i}{2}\bar{\psi}^{\DB
c}\gamma^\lambda\psi^d_\DB)f_{cd}{}^a{}_b\nonumber\\
&&+(Z^\pc_\DA D^\lambda\bar Z^{\DA\pd}-\frac{i}{2}\bar{\psi}^{B
\pc}\gamma^\lambda\psi^\pd_B)f_{\pc\pd}{}^a{}_b]\}
\nonumber\\
&&+ {\cal O}(Z^4).\label{4SusyOnA}
\end{eqnarray}
The last term ${\cal O}(Z^4)$, which is fourth order in the scalar
fields $Z$, vanishes by the FI (\ref{FI4}). The second line and the
third line must be the equations of motion for the gauge fields:
\begin{equation}\tilde{F}_{\mu\nu}{}^a{}_b=\varepsilon_{\mu\nu\lambda}
[(Z^c_AD^\lambda \bar Z^{Ad}-\frac{i}{2}\bar{\psi}^{\DB
c}\gamma^\lambda\psi^d_\DB)f_{cd}{}^a{}_b +(Z^\pc_\DA D^\lambda \bar
Z^{\DA\pd}-\frac{i}{2}\bar{\psi}^{B
\pc}\gamma^\lambda\psi^\pd_B)f_{\pc\pd}{}^a{}_b], \end{equation}
while the first line remains:
\begin{eqnarray}
[\delta_1, \delta_2]\tilde{A}_\mu{}^a{}_b&=&
v^\nu\tilde{F}_{\nu\mu}{}^a{}_b -D_\mu\tilde\Lambda^{a}{}_b.
\end{eqnarray}
The first term is a covariant translation; the  second term is a
gauge transformation, as expected. Similarly, we have
\begin{eqnarray}
[\delta_1, \delta_2]\tilde{A}_\mu{}^\pa{}_\pb&=&
v^\nu\tilde{F}_{\nu\mu}{}^\pa{}_\pb -D_\mu\tilde\Lambda^{\pa}{}_\pb,
\end{eqnarray}
and
\begin{equation}\tilde{F}_{\mu\nu}{}^\pa{}_\pb=\varepsilon_{\mu\nu\lambda}
[(Z^\pc_\DA D^\lambda\bar Z^{\DA\pd}-\frac{i}{2}\bar{\psi}^{B
\pc}\gamma^\lambda\psi^\pd_B)f_{\pc\pd}{}^\pa{}_\pb +(Z^c_A
D^\lambda\bar Z^{Ad}-\frac{i}{2}\bar{\psi}^{\DB
c}\gamma^\lambda\psi^d_\DB)f_{cd}{}^\pa{}_\pb]. \end{equation}

Finally we examine the fermion supersymmetry transformation:
\begin{eqnarray}\label{4SusyOnPsi}
\nonumber [\delta_1,\delta_2]\psi^a_{\DA} &=& v^\mu D_\mu
\psi^a_{\DA} + \tilde{\Lambda}^a{}_{b}
\psi^b_{\DA}\\
\nonumber &&-\frac{i}{2}(\epsilon_1^{\dag \DC B}\epsilon_{2B\DA}
-\epsilon_2^{\dag \DC B}\epsilon_{1B\DA})E^a_{\DC}\\
 &&
-\frac{1}{2}v_\nu\gamma^\nu E^a_{\DA},
\end{eqnarray}
where
\begin{equation}
E^a_{\DA} = \gamma^\mu D_\mu\psi^a_{\DA}
+f_{cdb}{}^aZ^b_BZ^{Bc}\psi^d_\DA-f_{\pc\pd b}{}^aZ^\pc_\DA
Z^\pd_\DC\psi^{\DC b}+2f_{\pc\pd b}{}^aZ^b_BZ^\pc_\DA\psi^{B\pd}.
\end{equation}
In order to achieve the closure of the algebra, we must impose the
equations of motion for the fermionic fields:
\begin{equation}
E^a_{\DA}=0.
\end{equation}
As a result, only the first line of (\ref{4SusyOnPsi}) remains.
Similarly, we obtain
\begin{eqnarray}\label{4SusyOnPsi2}
\nonumber [\delta_1,\delta_2]\psi^\pa_{A} &=& v^\mu D_\mu
\psi^\pa_{A} + \tilde{\Lambda}^\pa{}_{\pb} \psi^\pb_{A},
\end{eqnarray}
and
\begin{equation}
0=E^\pa_{A} = \gamma^\mu D_\mu\psi^\pa_{A} +f_{\pc\pd\pb}{}^\pa
Z^\pb_\DB Z^{\DB\pc}\psi^\pd_A-f_{cd\pb}{}^\pa Z^c_A Z^d_C\psi^{C
\pb}+2f_{cd \pb}{}^\pa Z^\pb_\DB Z^c_A\psi^{\DB d}.
\end{equation}
One can derive all the equations of motion of as the
Euler-Lagrangian equations from the Lagrangian (\ref{LN4}).

\section{Three-algebras, Lie superalgebras and Embedding
Tensors}\label{TLE}
\subsection{Three-algebras and Lie superalgebras}\label{super3}
In this section, we will demonstrate that the symplectic 3-algebra
can be realized in terms of a super Lie algebra.

Recall that in Sec. \ref{N1to5}, we note that $f_{IJKL}$ can be
specified as $k_{mn}\tau^m_{IJ}\tau^n_{KL}$ (up to an unimportant
constant), i.e.
\begin{equation}\label{solution}
f_{IJKL}=k_{mn}\tau^m_{IJ}\tau^n_{KL},
\end{equation}
where the set of matrices $\tau^m_{IK}$ is in the fundamental
representation of $Sp(2L)$ or its subalgebra, and $k_{mn}$ is the
Killing-Cartan metric.

Further more, the constraint condition $f_{(IJK)L}=0$ implies that
$f_{(IJK)L}=k_{mn}\tau^m_{(IJ}\tau^n_{K)L}=0$. As GW pointed out
\cite{GaWi}, the constraint equation
$k_{mn}\tau^m_{(IJ}\tau^n_{K)L}=0$ can be solved in terms of the
Jacobi identity for following super Lie algebra: \footnote{This is
\emph{not} the $D=3$ super-Pioncare algebra.}
\e\label{SLie} &&[M^m, M^n]=C^{mn}{}_sM^s,\nonumber\\
&&[M^m, Q_I]=-\t^m_{IJ}\omega^{JK}Q_K,\nonumber\\
&&\{Q_I,Q_J\}=\t^m_{IJ}k_{mn}M^n.\ee
Namely, the $QQQ$ Jacobi identity
\e [\{Q_I,Q_J\}, Q_K]+[\{Q_J,Q_K\}, Q_I]+[\{Q_K,Q_I\}, Q_J]=0\ee
is equivalent to the constraint equation
$k_{mn}\tau^m_{(IJ}\tau^n_{K)L}=0$. Therefore GW's approach
suggests that the symplectic 3-algebra can be realized in terms of
the super Lie algebra (\ref{SLie}), if we think of the 3-algebra
generator $T_I$ as the fermionic generator $Q_I$. Comparing the
3-bracket $[T_I, T_J; T_K]=f_{IJK}{}^{L}T_l$ with
\begin{equation}\label{dcommutator}
[\{Q_I,Q_J\}, Q_K]=k_{mn}\tau^m_{IJ}\tau^n_{K}{}^LQ_L,
\end{equation}
and taking account of (\ref{solution}), we see that the 3-bracket
may be realized in terms of the double graded commutator
\begin{equation}
[T_I, T_J; T_K]\doteq[\{Q_I,Q_J\}, Q_K].
\end{equation}
Here the RHS is also obviously symmetric in $IJ$. It is instructive
to examine the FI (\ref{FI}) with the 3-brackets replaced by the
double graded commutators:
\begin{eqnarray}
&&[\{Q_I,Q_J\},[\{Q_M,Q_N\},Q_K]]\\&=&[\{[\{Q_I,Q_J\},Q_M],Q_N\},
Q_K]+[\{Q_M,[\{Q_I,Q_J\},Q_N]\},
Q_K]+[\{Q_M,Q_N\},[\{Q_I,Q_J\},Q_K]]\nonumber.
\end{eqnarray}
By using the super Lie algebra (\ref{SLie}), we obtain
\e \t^m_{IJ}\t^n_{MN}([M_n,[M_m,Q_K]]-[M_m,[M_n,Q_K]]+[[M_m,
M_n],Q_K])=0,\ee
which is equivalent to the $MMQ$ Jacobi identity of the super Lie
algebra (\ref{SLie}). It is not difficult to prove that
$k_{mn}\tau^m_{IJ}\tau^n_{KL}$ also enjoy the symmetry properties
(\ref{SymmeOfF}). So indeed the symplectic 3-algebra can be realized
in terms of the super Lie algebra. Now recall the component
formulism of the basic definition of the global transformation
\begin{equation}
\d_{\tilde\Lambda}X^K=\Lambda^{IJ}f_{IJ}{}^K{}_LX^L.
\end{equation}
Replacing $f_{IJ}{}^K{}_L$ by $k_{mn}\t^m_{IJ}\t^n{}^K{}_L$ gives
\begin{equation}\label{lietran}
\d_{\tilde\Lambda}X^K=\Lambda^{IJ}k_{mn}\t^m_{IJ}\t^n{}^K{}_LX^L.
\end{equation}
From the ordinary Lie group point of view, this is a transformation with
parameters $\Lambda^{IJ}k_{mn}\t^m_{IJ}$ and generators
$\t^n{}^K{}_L$. On the other hand, the second equation of
(\ref{SLie}) indicates that the fermionic generators furnish a
representation of the bosonic part of the super Lie algebra
(\ref{SLie}), i.e. the matrix $\t^m_{IJ}$ is a quaternion
representation of $M^m$. Therefore, the gauge group generated by the
3-algebra can be determined as follows: its Lie algebra is just the
bosonic part of the super Lie algebra (\ref{SLie}), which must be
$Sp(2L)$ or its sub-algebras. The representation of the matter
fields is determined by the fermionic generators of the super Lie
algebra (\ref{SLie}).

For a more mathematical approach, see Ref. \cite{MFM:Aug09, Jose,
Jakob}, in which the relations between the 3-algebras and Lie
superalgebras are discussed by using Lie algebra representation
theories.

\subsection{Three-algebras and Lie Algebras}\label{seclie}
It is less obvious that one can also prove that (\ref{solution})
is an explicit solution of the FI (\ref{FFI}) by using the $QQM$
Jacobi identity of the super Lie algebra, which reads
\e [\{Q_I,Q_J\}, M^m]-\{[Q_J,M^m],Q_I\}+\{[M^m,Q_I],Q_J\}=0. \ee
After a short algebra we obtain
\e \t^n_{IJ}k_{np}[M^p,
M^m]-\t^{mK}{}_J\t^n_{KI}k_{np}M^p-\t^{mK}{}_I\t^n_{KJ}k_{np}M^p=0.\ee
Since the matrix $\t^m_{IJ}$ is a representation of $M^m$, the above
equation implies
\e \t^n_{IJ}k_{np}[\t^p,
\t^m]_{MN}-\t^{mK}{}_J\t^n_{KI}k_{np}\t^p_{MN}-\t^{mK}{}_I\t^n_{KJ}k_{np}\t^p_{MN}=0,\ee
where \begin{equation}[\t^p,
\t^m]_{MN}=\t^p_{MO}\t^{mO}{}_N-\t^m_{MO}\t^{pO}{}_N.
\end{equation}
Multiplying both sides by $k_{mq}\t^q_{KL}$ gives
\begin{equation}\label{FItensor} k_{np}\t^n_{IJ}k_{mq}\t^q_{KL}[\t^p,
\t^m]_{MN}-k_{mq}\t^q_{KL}\t^{mK}{}_J\t^n_{KI}
k_{np}\t^p_{MN}-k_{mq}\t^q_{KL}\t^{mK}{}_I\t^n_{KJ}k_{np}\t^p_{MN}=0.
\end{equation}
Rearranging the above equation verifies explicitly that
(\ref{solution}) satisfies the FI (\ref{FFI}). Application of the
commutator
\e\label{commutator1}[\t^m, \t^n]_{IJ}=C^{mn}{}_p\t^p_{IJ}\ee
to Eq. (\ref{FItensor}) gives
\begin{equation}
\label{C}
(k_{np}k_{qm}C^{pm}{}_s+k_{qm}k_{sp}C^{pm}{}_n)\t^n_{IJ}\t^q_{KL}\t^s_{MN}=0.
\end{equation}
Here the equation in the bracket is simply the statement that the
structure constants 
\begin{equation}\label{tanti}
\tilde C_{nqs}=k_{np}k_{qm}C^{pm}{}_s
\end{equation}
are totally antisymmetric if the three adjoint indices $nqs$ are on
equal footing. Note that $k_{mn}$ is an invariant bilinear form on
the bosonic subalgebra, since Eq. (\ref{C}) or (\ref{tanti}) also
implies
\begin{equation}\label{commutator2}
\quad [k,C^m]=0.
\end{equation}
Here the matrices $(C^m)^p{}_n=C^{mp}{}_n$ furnish the usual adjoint
representation of the bosonic subalgebra. In this way, we see that
the FI of the 3-algebra can be converted into two ordinary
commutators (\ref{commutator1}) and (\ref{commutator2}) (this is
first discovered in the second paper of Ref. \cite{Gustavsson} with
a different approach).

Eq. (\ref{lietran}) indicates that
$f_{IJKL}=k_{mn}\t^m_{IJ}\t^n_{KL}$ also furnish a quaternion
representation of the bosonic subalgebra. In fact, if we write
$f_{IJKL}$ as $(f_{IJ})_{KL}$, then $f_{IJ}$ is a set of matrices,
and corresponding matrix elements are $(f_{IJ})_{KL}$. If
$\t^n_{KL}$ furnish a quaternion of representation of $M^n$, then
$(f_{IJ})_{KL}$ furnish a quaternion representation of
$M_{IJ}=k_{mn}\t^m_{IJ}M^n$, since the operator $M_{IJ}$ is a linear
combination of $M^n$. With this understanding, we are able to
re-write the FI (\ref{FFI}) as a commutator
\e\label{commutator} [f_{IJ},
f_{KL}]_{MN}&=&C_{IJ,KL}{}^{OP}(f_{OP})_{MN}\nonumber\\
&=&(f_{IJK}{}^O\d_L^P+f_{IJL}{}^O\d_K^P)(f_{OP})_{MN}\nonumber\\
&=&-[f_{IJ},f_{MN}]_{KL}. \ee
The third equation says that the quantity $[f_{IJ}, f_{KL}]_{MN}$
are totally antisymmetric in the 3 pairs of indices. Eq.
(\ref{commutator}) is equivalent to Eq. (\ref{commutator1}). Also,
the matrices $(f_{IJ})_{KL}$ satisfy the conventional Jacobi
identity as a result of the $MMM$ Jacobi identity of the
superalgebra of (\ref{SLie}). We now must check whether $\tilde
C_{IJ,KL,MN}=k_{MN,OP}C_{IJ,KL}{}^{OP}$ are totally antisymmetric
or not. To be consistent with the transformation
$M_{IJ}=k_{mn}\t^m_{IJ}M^n$, we must transform the Killing-Cartan
metric $k^{mn}$ as
\e\label{kc} k^{mn}\rightarrow
k_{IJ,KL}=k_{qm}\t^q_{IJ}k_{pn}\t^p_{KL}k^{mn}=k_{mn}\t^m_{IJ}\t^n_{KL}=f_{IJKL}.\ee
Namely the structure constants $f_{IJKL}$ also play a role of the
Killing-Cartan metric $k_{IJ,KL}$. So we must use $f_{MNOP}$ to
lower the $OP$ indices of $C_{IJ,KL}{}^{OP}$: \footnote{This is a
comment by E. Witten, quoted in the second paper of Ref.
\cite{Gustavsson}.}
\e\label{strofcm} \tilde C_{IJ,KL,MN}&=&f_{MNOP}C_{IJ,KL}{}^{OP}\nonumber\\
&=& [f_{MN},f_{IJ}]_{KL}.\ee
By the third equation of (\ref{commutator}), the structure constants
$\tilde C_{IJ,KL,MN}$ are indeed totally antisymmetric in the 3
pairs of indices. Therefore Eq. (\ref{commutator2}) now takes the
following form
\e \label{totantisym}[f,C_{IJ}]=0\quad {\rm or}\quad
[f_{MN},f_{IJ}]_{KL}+[f_{KL},f_{IJ}]_{MN}=0,\ee
which is nothing but the third equation of Eq. (\ref{commutator}).
Namely both Eq. (\ref{commutator1}) and Eq. (\ref{commutator2}) can
be written as the third equation of Eq. (\ref{commutator}), if we
express everything in terms of the 3-algebra structure constants
$f_{IJKL}$.

Note that we use $k_{mn}$ to lower an adjoint index, while use
$\omega_{IJ}$ to lower a fundamental index. If Eq. (\ref{solution})
holds, then Eq. (\ref{DeltaOmg}) implies a compatible condition
between $k_{mn}$ and $\omega_{IJ}$. Eq. (\ref{DeltaOmg}) is
equivalent to
$k_{nm}\t^{mK}_I\omega_{KJ}+k_{nm}\t^{mK}_J\omega_{IK}=0$, i.e.
\begin{equation}
\tilde\t_{nIJ}-k_{nm}\omega_{IK}\t^{mK}_J=0,
\end{equation}
where $\tilde\t_{nIJ}\equiv k_{nm}\t^m_{IJ}$.

\subsection{Three-Algebras and Embedding Tensors}\label{embed}
In Ref. \cite{Bergshoeff:2008cz, Bergshoeff}, the authors derive
some extended superconformal gauge theories by taking a conformal
limit of $D=3$ gauged supergravity theories. In their approach, the
embedding tensor plays a crucial role. By definition, the embedding
tensor $\theta_{mn}=\theta_{nm}$ acts as a projector
\cite{Bergshoeff}:
\begin{equation}
D_\mu=\partial_\mu-A^m_\mu\theta_{mn}t^n,
\end{equation}
where $t^n$ is a set of independent generators. The above equation
says that $\theta_{mn}$ projects $t^n$ onto another set of
generators $\tilde t_m=\theta_{mn}t^n$, whose symmetries are gauged.
Let us now consider the commutator
\e [\tilde t_m,\tilde t_n]=\theta_{mp}\theta_{ns}C^{ps}{}_qt^q.\ee
Since we expect that $[\tilde t_m,\tilde t_n]=\tilde
C_{mn}{}^r\tilde t_r$, we must set
\e\label{pro} \theta_{mp}\theta_{ns}C^{ps}{}_q=\tilde
C_{mn}{}^r\theta_{rq}.\ee
It is necessary to examine the Jacobi identity
\e\label{jacobi} &&[[\tilde t_m,\tilde t_n],\tilde t_p]+[[\tilde
t_n,\tilde
t_p],\tilde t_m]+[[\tilde t_p,\tilde t_m],\tilde t_n]\nonumber\\
&=&(\tilde C_{mn}{}^s\tilde C_{sp}{}^r+\tilde C_{np}{}^s\tilde
C_{sm}{}^r+\tilde C_{pm}{}^s\tilde C_{sn}{}^r)\theta_{rq}t^q\nonumber\\
&=&(C^{lq}{}_rC^{rs}{}_t+C^{qs}{}_rC^{rl}{}_t+C^{sl}{}_rC^{rq}{}_t)
\theta_{ml}\theta_{nq}\theta_{ps}t^t=0.\ee
In the last line we have used (\ref{pro}). The last line is nothing
but the Jacobi identity satisfied by $C^{mn}{}_p$. So Eq.
(\ref{jacobi}) is indeed the desired result. To construct a physical
theory, the embedding tensor is required to be invariant under the
transformations which are gauged. Since the embedding tensor
$\theta_{mn}$ carries two adjoint indices, we have to set
\begin{equation}
\tilde C_{nq}{}^r\theta_{rs}+\tilde C_{ns}{}^r\theta_{qr}=0.
\end{equation}
Taking account of (\ref{pro}), the above equation is equivalent to
\begin{equation}\label{C2}
\theta_{np}\theta_{qm}C^{pm}{}_s+\theta_{np}\theta_{sm}C^{pm}{}_q=0.
\end{equation}
This quadratic constraint takes the same form for all extended
supergravity theories. We will focus on the $\CN=5$ case. If we
represent the adjoint index $m$ as a pair of fundamental indices
$IJ$, the embedding tensor becomes $\theta_{IJ,KL}$, satisfying the
same symmetry properties as $f_{IJKL}$ do (see (\ref{SymmeOfF}))
\cite{Bergshoeff:2008cz}. To construct $\CN=5$ supergravity
theories, the embedding tensor is required to satisfy the linear
constraint:
\begin{equation}\label{linear}
\theta_{(IJ,K)L}=0,
\end{equation}
and the structure constants in (\ref{C2}) are required to be those
of $Sp(2L)$ \cite{Bergshoeff:2008cz}. We observe that if one
identifies the embedding tensor $\theta_{mn}$ with the
Killing-Cartan metric $k_{mn}$, Eq. (\ref{C2}) is precisely the
same as Eq. (\ref{C}), which is the FI satisfied by the 3-algebra
structure constants $f_{IJKL}=k_{mn}\t^m_{IJ}\t^n_{KL}$. Recall
that $f_{IJKL}$ also play the role of the Killing-Cartan metric
(see Sec. \ref{seclie}). So identifying the embedding tensor
with the Killing-Cartan metric is equivalent to identifying the
embedding tensor with the 3-algebra structure constants. With this
identification, Eq. (\ref{linear}) is also solved since it is
nothing but $f_{(IJK)L}=0$. We are therefore led to the conclusion
that $f_{IJKL}$ also play the role of the embedding tensor. It is
straightforward to generalize the discussion of this section to
the cases with other values of $\CN$.

In summary, if we realize the symplectic 3-algebra in terms of the
superalgebra (\ref{SLie}), we find that
$f_{IJKL}=k_{mn}\t^m_{IJ}\t^n_{KL}$ play four roles simultaneously:
\begin{itemize}
\item $f_{IJKL}$ are the structure constants of the symplectic 3-algebra
or the double graded commutator (\ref{dcommutator});
\item $f_{IJKL}$ furnish a quaternion representation
of the bosonic part of the superalgebra;
\item $f_{IJKL}$ play the role of the Killing-Cartan
metric;
\item $f_{IJKL}$ are the components of the embedding tensor
used to construct the $D=3$ extended supergravity theories.
\end{itemize}

\section{\CN=4, 5 Theories in Terms of Lie Algebras}\label{LieN4N5}
The $\CN=4, 5$ theories in Sec. \ref{N5} and \ref{secN4} are
constructed in terms of 3-algebras. After the discussions of the
last section, we are ready to derive their ordinary Lie Algebra
constructions by the solution (\ref{solution}).

\subsection{$\CN=5$ Theories in Terms of Lie Algebras}
With the solution
\begin{equation}\label{soln}
f_{IJKL}=k_{mn}\t^m_{IJ}\t^n_{KL},\quad [\t^m,
\t^n]_{IJ}=C^{mn}{}_p\t^p_{IJ},
\end{equation}
the gauge field becomes
\begin{equation}\label{gauge}
\tilde
A_\mu{}^K{}_L=A_\mu^{IJ}f_{IJ}{}^K{}_L=A_\mu^{IJ}\t^m_{IJ}k_{mn}\t^{nK}{}_{L}\equiv
A^m_\mu k_{mn}\t^{nK}{}_{L}.
\end{equation}
Following Ref. \cite{GaWi}, we define the `momentum map' and
`current ' operator as follows
\e \mu^m_{AB}\equiv \t^m_{IJ}Z^I_AZ^J_B, \quad
j^m_{AB}\equiv\t^m_{IJ}Z^I_A\p^J_B.\ee
Here $A=1,\ldots, 4$ is the fundamental index of the R-symmetry
group $Sp(4)$. Substituting the (\ref{soln}) and (\ref{gauge}) into
the Lagrangian (\ref{GeneN5Lagran}) gives
\begin{eqnarray}\label{5lagran}\nonumber
{\cal L}&=&\frac{1}{2}(-D_\mu\bar{Z}^A_ID^\mu
Z^I_A+i\bar{\psi}^A_I\gamma_\mu
D^\mu\psi^I_A)-\frac{i}{2}\omega^{AB}\omega^{CD}k_{mn}(j^m_{AC}j^n_{BD}-
2j^m_{AC}j^n_{DB})\nonumber\\
&&+\frac{1}{2}\epsilon^{\mu\nu\lambda}(k_{mn}A_\mu^m\partial_\nu
A_\lambda^n+\frac{1}{3}\tilde C_{mnp}A_\mu^mA_\nu^nA_\lambda^p)\\
&&+\frac{1}{30}\tilde C_{mnp}\mu^{mA}{}_B\mu^{nB}{}_C\mu^{pC}{}_A
+\frac{3}{20}k_{mp}k_{ns}(\t^m\t^n)_{IJ}Z^{AI}Z^J_A\mu^{pB}{}_C\mu^{sC}{}_B.\nonumber
\end{eqnarray}
Similarly, with the solution ({\ref{soln}), the supersymmetry
transformation law becomes
\begin{eqnarray}\label{n5susy}\nonumber
\delta Z^I_A&=&i\epsilon_A{}^{B}\psi^I_{B},\nonumber\\
\delta\psi^I_{A}&=&\gamma^{\mu}D_\mu Z^I_B\epsilon^B{}_{A}
+\frac{1}{3}k_{mn}\t^{mI}{}_{J}\omega^{BC}Z^J_B\mu^n_{CD}\epsilon^D{}_{A}
-\frac{2}{3}k_{mn}\t^{mI}{}_{J}\omega^{BD}Z^J_C\mu^n_{DA}\epsilon^C{}_{B},
\nonumber\\
\delta A^m_\mu&=& i\epsilon^{AB}\gamma_\mu j^m_{AB}.
\end{eqnarray}
Here the parameter $\ep_A{}^B$ obeys the traceless condition and the
reality condition (\ref{SusyPara5}). The $\CN=5$ Lagrangian
(\ref{5lagran}) and supersymmetry transformation law (\ref{n5susy})
are in agreement with those given in Ref. \cite{Hosomichi:2008jb},
which were derived directly in terms of ordinary Lie algebra.

In section (\ref{super3}), we demonstrate that if the structure
constants of the 3-algebra are specified as (\ref{soln}), then the
Lie algebra of the gauge group generated by the 3-algebra is just
the bosonic part of the superalgebra (\ref{SLie}). The following
classical super-Liealgebras:
\begin{equation}\label{listsa}
U(M|N),\quad OSp(M|2N),\quad  OSp(2|2N),\quad F(4),\quad G(3),\quad
D(2|1;\a),
\end{equation}
(with $\a$ a continuous parameter) are of the same form as that of
the superalgebra (\ref{SLie}). Therefore their bosonic parts can
be selected to be the Lie algebras of the gauge groups of the
$\CN=5$ theories. Especially, if we choose the $U(M|N)$ or
$OSp(2|2N)$, whose bosonic part is in the two conjugate
representations ($R\oplus\bar R$), then the supersymmetry will get
enhanced to $\CN=6$ \cite{Hosomichi:2008jb}. In the case of $OSp(M|2N)$,
the theory has been
conjectured to be the dual gauge theory of M2-branes in orbifold
$\textbf{C}^4/\hat{\textbf{D}}_k$, with
$\hat{\textbf{D}}_k$ the binary dihedral group
\cite{Hosomichi:2008jb, Aharony:2008gk}. The gravity dual of this theory has
been investigated in Ref. \cite{Aharony:2008gk}.

\subsection{$\CN=4$ GW Theory in Terms of Lie
Algebras}\label{secgw}

Here we consider only the $\CN=4$ GW theory without the `twisted'
hyper multiplets, i.e., setting $\Phi^\pa_\DA=0$. Then with the
solution for structure constants of the 3-algebra given by
\begin{eqnarray}\label{n4ten}
&&f_{abcd}=k_{mn}\t^m_{ab}\t^n_{cd}, \quad [\t^m,
\t^n]_{ab}=C^{mn}{}_p\t^p_{ab},
\end{eqnarray}
which satisfy the FI's as well as appropriate constraints and
symmetry conditions, the gauge fields of the GW theory become
\begin{eqnarray}\label{GWgauge}
\tilde A_\mu{}^c{}_d&=&A_\mu^{ab}f_{ab}{}^c{}_d
=A_\mu^{ab}\t^m_{ab}k_{mn}\t^{nc}{}_d \equiv
A_\mu^mk_{mn}\t^{nc}{}_d.
\end{eqnarray}
Following Ref. \cite{GaWi}, we define the `momentum map' and
`current ' operators as follows
\e\label{mmmp} \mu^m_{AB}\equiv \t^m_{ab}Z^a_AZ^b_B, \quad
j^m_{A\DB}\equiv\t^m_{ab}Z^a_A\p^b_\DB.\ee
With Eqs (\ref{n4ten}) $\sim$ (\ref{mmmp}), Eqs. (\ref{LN4GW}) and
(\ref{SUSY4GW}) become the Lagrangian and the supersymmetry law of
the GW theory in Ref. \cite{GaWi}, respectively:
\e\label{gwlgr}
\CL&=&\frac{1}{2}\epsilon^{\mu\nu\lambda}(k_{mn}A_\mu^m\partial_\nu
A_\lambda^n+\frac{1}{3}\tilde
C_{mnp}A_\mu^mA_\nu^nA_\lambda^p)+\frac{1}{2}(-D_\mu\bar{Z}^A_aD^\mu
Z^a_A+i\bp^\DA_a\g^\mu
D_\mu\p^a_\DA)\nonumber\\
&&-\frac{i}{2}k_{mn}j^m_{A\DB}j^{nA\DB} -\frac{1}{24}\tilde
C_{mnp}\mu^{mA}{}_B\mu^{nB}{}_C\mu^{pC}{}_A , \ee
with $\tilde C_{mnp}=k_{mr}k_{ns} C^{rs}{}_p$ and
\e \label{gwsusy}&&\delta Z^a_A=i\ep_A{}^\DA\p^a_\DA,\nonumber\\
&&\delta\p^a_\DA=-\g^\mu D_\mu
Z^a_B\ep^\dag_\DA{}^B-\frac{1}{3}k_{mn}\t^{ma}{}_{b}Z^b_B\mu^{
nB}{}_C\ep^\dag_\DA{}^C,\nonumber\\
&&\delta A_\mu^m=i\ep^{A\DB}\g_\mu j^m_{A\DB}. \ee
Since we derived the GW theory by decomposing the $\CN=5$ theory
and setting the twisted mulitplets to zero, the classical
superalgebras, that are used to realize the 3-algebra, must be
the same as those used in the $\CN=5$ case, i.e.
\begin{equation}\label{listsa2}
U(M|N),\quad OSp(M|2N),\quad  OSp(2|2N),\quad F(4),\quad G(3),\quad
D(2|1;\a).
\end{equation}
Indeed, they are of the same form as that of the superalgebra
(\ref{SLie}). Therefore their bosonic parts can be selected to be
the Lie algebras of the gauge groups of the GW theory; and the
corresponding representations are determined by the fermionic
generators.

For the cases \cite{HosomichiJD,Hosomichi:2008jb} with both
un-twisted and twisted hyper-multiplets, a pair of super Lie
algebras are needed, which were discussed in a representation theory
approach in Ref. \cite{MFM:Aug09}. Since the situation is much more
complicated, we leave the presentation of these cases in terms of
ordinary Lie algebras, as well as their generalizations, within our
superspace and super Lie algebra approach to a subsequent paper
\cite{ChenWunew}.

\section{Conclusions}
In this paper, we have combined the symplectic 3-algebra with the
superspace formalism by letting the matter superfields take values
in the symplectic 3-algebra. Based on the 3-algebra, we then have
constructed the general $\CN=5$ CMS theory by enhancing the
$\CN=1$ supersymmetry to $\CN=5$. The $\CN=5$ Lagrangian is same
as the one derived with an on-shell approach \cite{Chen2}.

We have constructed the general $\CN=4$ CSM theory by decomposing
one $\CN=5$ hypermultiplet into a $\CN=4$ un-twisted hypermultiplet and
a $\CN=4$ twisted hypermultiplet, and then proposing a new
superpotential. In deriving the general $\CN=4$ CSM theory, we
have also decomposed the set of 3-algebra generators into two sets
of 3-algebra generators. As a result, both the FI's and 3-brackets
are decomposed into 4 sets. The resulting general $\CN=4$ CSM
theory is a quiver gauge theory based on the 3-algebra. We have
also examined the closure of the $\CN=4$ algebra.

We then have realized the symplectic 3-algebra in terms of the
super Lie algebra (\ref{SLie}). The 3-bracket is realized in terms
of a double graded bracket: $[T_I,T_J; T_K]\doteq [\{Q_I, Q_J\},
Q_K]$, where $Q_I$ are the fermionic generators; the structure
constants of the 3-algebra are just the structure constants of the
double graded bracket, i.e., $f_{IJKL}=k_{mn}\t^m_{IJ}\t^n_{KL}$.
The fundamental identity of the 3-algebra is equivalent to the
$MMQ$ Jacobi identity of the super Lie algebra, where $M$'s are
the bosonic generators in the super Lie algebra. The linear
constraint equation $f_{(IJK)L}=0$, required by the enhancement of
the supersymmetry, is equivalent to the $QQQ$ Jacobi identity.

We have also analyzed the relations between the symplectic
3-algebra and the ordinary Lie algebra. The fundamental identity
of 3-algebra can be solved in terms of a tensor product:
$f_{IJKL}=k_{mn}\t^m_{IJ}\t^n_{KL}$. We have proved that the
structure constants $f_{IJKL}$ furnish a quaternion representation
of the bosonic part of the super Lie algebra (\ref{SLie}), and
$f_{IJKL}$ also play a role of Killing-Cartan metric. We found
that the FI of the 3-algebra can be converted into an ordinary
commutator (\ref{commutator}); the structure constants of the
commutator are (\ref{strofcm}). The FI of the 3-algebra can be
understood as the statement that the structure constants of the
commutator (\ref{strofcm}) are total antisymmetric (see Eqs.
(\ref{totantisym})).

We have proved that the components of an embedding tensor
\cite{Bergshoeff:2008cz, Bergshoeff}, used to construct the $D=3$
extended supergravity theories, are just the structure constants
of the 3-algebra. Hence the concepts and techniques of the
3-algebra may be used to construct new $D=3$ extended supergravity
theories.

The general ${\cal N}=5$ CSM theories and the $\CN=4$ GW CSM
theories in terms of ordinary Lie algebras are rederived,
respectively, in our superspace approach. The presentation of
general $\CN=4$ CSM theories is left for a subsequent paper
\cite{ChenWunew}. In this way, we have been able to derive all known
${\cal N}=4,5$ superconformal Chern-Simons matter theories, as well
as some new $\CN=4$ quiver gauge theories (to be presented in
\cite{ChenWunew}). Thus our superspace formulation for the
super-Lie-algebra realization of symplectic 3-algebras provides a
unified treatment of all known $\CN=4,5,6,8$ CSM theories, including
new examples of $\CN=4$ quiver gauge theories as well.

The extended ($\CN\geq4$) CSM theories can be also constructed by using $\CN=2,3$ superspace formulations in an ordinary \emph{Lie (2-)algebra} approach \cite{Benna, N3sps1, Buchbinder2}. The $\CN=2,3$ superspace formulations are more restrictive than the $\CN=1$ formulation; hence the calculations may be simplified, and it may be easier to enhance the supersymmetry from $\CN=2$ or $\CN=3$ to $\CN=6,8$. It would be nice to construct the extended CSM theories by using the $\CN=2,3$ superspace formulations in a  \emph{3-algebra} approach.


\section{Acknowledgement}

YSW is supported in part by the US NSF through Grant No.
PHY-0756958. He also thanks for the Department of Physics, Fudan
University for warm hospitality during the completion stage of
this work.

\appendix

\section{Conventions and Useful Identities}\label{Identities}

\subsection{Spinor Algebra}
In $1+2$ dimensions, the gamma matrices are defined as
\begin{equation}
(\gamma_{\mu})_{\alpha}{}^\gamma(\gamma_{\nu})_{\gamma}{}^\beta+
(\gamma_{\nu})_{\alpha}{}^\gamma(\gamma_{\mu})_{\gamma}{}^\beta=
2\eta_{\mu\nu}\delta_{\alpha}{}^\beta.
\end{equation} For the metric we
use the $(-,+,+)$ convention. The gamma matrices in the Majorana
representation can be defined in terms of Pauli matrices:
$(\gamma_{\mu})_{\alpha}{}^\beta=(i\sigma_2, \sigma_1, \sigma_3)$,
satisfying the important identity
\begin{equation}
(\gamma_{\mu})_{\alpha}{}^\gamma(\gamma_{\nu})_{\gamma}{}^\beta
=\eta_{\mu\nu}\delta_{\alpha}{}^\beta+\varepsilon_{\mu\nu\lambda}(\gamma^{\lambda})_{\alpha}{}^\beta.
\end{equation}
We also define
$\varepsilon^{\mu\nu\lambda}=-\varepsilon_{\mu\nu\lambda}$. So
$\varepsilon_{\mu\nu\lambda}\varepsilon^{\rho\nu\lambda} =
-2\delta_\mu{}^\rho$. We raise and lower spinor indices with an
antisymmetric matrix
$\epsilon_{\alpha\beta}=-\epsilon^{\alpha\beta}$, with
$\epsilon_{12}=-1$. For example,
$\psi^\alpha=\epsilon^{\alpha\beta}\psi_\beta$ and
$\gamma^\mu_{\alpha\beta}=\epsilon_{\beta\gamma}(\gamma^\mu)_\alpha{}^\gamma
$, where $\psi_\beta$ is a Majorana spinor. Notice that
$\gamma^\mu_{\alpha\beta}=(\mathbbm{l}, -\sigma^3, \sigma^1)$ are
symmetric in $\alpha\beta$. A vector can be represented by a
symmetric bispinor and vice versa:
\begin{equation}
A_{\alpha\beta}=A_\mu\gamma^\mu_{\alpha\beta},\quad\quad A_\mu=-\frac{1}{2}\gamma^{\alpha\beta}_\mu A_{\alpha\beta}.
\end{equation}
We use the following spinor summation convention:
\begin{equation}
\psi\chi=\psi^\alpha\chi_\alpha,\quad\quad
\psi\gamma_\mu\chi=\psi^\alpha(\gamma_{\mu})_{\alpha}{}^\beta\chi_\beta,
\end{equation}
where $\psi$ and $\chi$ are anti-commuting Majorana spinors. In
$1+2$ dimensions the Fierz transformations are
\begin{eqnarray}
(\lambda\chi)\psi &=& -\frac{1}{2}(\lambda\psi)\chi -\frac{1}{2}
(\lambda\gamma_\nu\psi)\gamma^\nu\chi,\\
(\psi_1\psi_2)(\psi_3\psi_4)&=&(\psi_1\psi_2)(\psi_4\psi_3)=-\frac{1}{2}(\psi_1\psi_3)(\psi_4\psi_2)-
\frac{1}{2}(\psi_1\gamma_\nu\psi_3)(\psi_4\gamma^\nu\psi_2),\nonumber\\
(\psi_1\gamma_\mu\psi_2)(\psi_3\psi_4)&=&-\frac{1}{2}(\psi_1\gamma_\mu\psi_3)(\psi_4\psi_2)-
\frac{1}{2}(\psi_1\psi_3)(\psi_4\gamma_\mu\psi_2)+\frac{1}{2}\varepsilon_{\mu\nu\lambda}
(\psi_1\gamma^\nu\psi_3)(\psi_4\gamma^\lambda\psi_2)\nonumber.
\end{eqnarray}

\subsection{The ${\cal N}=1$ Superspace}
In this subsection, we mainly follow the conventions of Ref.
\cite{Hosomichi:2008jb}. We denote the superspace coordinates as
$\theta^\alpha$. A real scalar superfield $\Phi$ can be expanded as
\begin{equation}
\Phi=\phi+i\theta\psi-\frac{i}{2}\theta^2F,
\end{equation}
where $\theta$ and $\psi$ are Majorana spinors. The superalgebra
\begin{equation}
\{Q_\alpha, Q_\beta\}=-2\gamma^\mu_{\alpha\beta}P_\mu
\end{equation}
can be realized in terms of superspace derivatives:
\begin{eqnarray}
Q_\alpha=i\partial_\alpha+\theta^\beta\partial_{\beta\alpha}.
\end{eqnarray}
The super-covariant derivative must anti-commute with $Q_\alpha$; it
takes the following form:
\begin{eqnarray}\label{SCD}
 \mathscr{D}_\alpha=\partial_\alpha+i\theta^\beta\partial_{\beta\alpha}.
\end{eqnarray}
The supersymmetry transformation of $\Phi$ is defined as
\begin{eqnarray}
\delta\Phi=-i\epsilon^\alpha Q_\alpha\Phi\equiv
\delta\phi+i\theta\delta\psi-\frac{i}{2}\theta^2\delta F.
\end{eqnarray}
Equating powers of $\theta^\alpha$ gives the supersymmetry
transformations of the component fields:
\begin{eqnarray}
\delta\phi&=& i\epsilon^\alpha\psi_\alpha,\\
\delta\psi_\alpha&=&-\partial_\alpha{}^\beta\phi\epsilon_\beta-F\epsilon_\alpha,\\
\delta F&=&i\epsilon^\alpha\partial_\alpha{}^\beta\psi_\beta.
\end{eqnarray}
In the Wess-Zumino gauge, the superconnection becomes
\begin{eqnarray}
\Gamma_\alpha&=&i\theta^\beta A_{\alpha\beta}
+\theta^2\chi_{\alpha},
\end{eqnarray}
and the supersymmetry transformations for the component fields are
\begin{eqnarray}
\delta A_\mu&=&-i\epsilon^\alpha (\gamma_\mu)_\alpha{}^\beta\chi_\beta,\\
\delta\chi_\alpha&=&-\frac{1}{2}F_{\mu\nu}(\gamma^{\mu\nu})_\alpha{}^\beta\epsilon_\beta.
\end{eqnarray}
The Berezin integral is defined as
\begin{equation}
\int d^2\theta\theta^2=-4.
\end{equation}
The superpotential is given by
\begin{eqnarray}
\CL_W=\frac{i}{2}\int d^2\theta
W(\Phi)=-\frac{i}{2}W^{\prime\prime}(\phi)\psi^2-W^{\prime}(\phi)F.
\end{eqnarray}

\subsection{$SU(2)\times SU(2)$ Identities}\label{SO4}
We define the 4 sigma matrices as
\begin{equation}\label{pulim}
\sigma^a{}_A{}^{\dot{B}}=(\sigma^1,\sigma^2,\sigma^3,i\mathbbm{l}),
\end{equation}
by which one can establish a connection between the $SU(2)\times
SU(2)$ and $SO(4)$ group. These sigma matrices satisfy the following
Clifford algebra:
\begin{eqnarray}
\sigma^a{}_{A}{}^{\dot{C}}\sigma^{b\dag}{}_{\dot{C}}{}^B+
\sigma^b{}_{A}{}^{\dot{C}}\sigma^{a\dag}{}_{\dot{C}}{}^B=2\delta^{ab}\delta_A{}^B,\\
\sigma^{a\dag}{}_{\dot{A}}{}^{C}\sigma^{b}{}_{C}{}^{\dot{B}}+
\sigma^{b\dag}{}_{\dot{A}}{}^{C}\sigma^{a}{}_{C}{}^{\dot{B}}=2\delta^{ab}\delta_{\dot{A}}{}^{\dot{B}}.
\end{eqnarray}
We use antisymmetric matrices
\begin{eqnarray}
\epsilon_{AB}=-\epsilon^{AB}=\begin{pmatrix} 0&-1 \\ 1&0
\end{pmatrix}\;
\;\;{\rm and}\;\;\;
\epsilon_{\dot{A}\dot{B}}=-\epsilon^{\dot{A}\dot{B}}=\begin{pmatrix}
0&1 \\-1& 0
\end{pmatrix}
\end{eqnarray}
to raise or lower un-dotted and dotted indices, respectively. For
example,
$\sigma^{a\dag\dot{A}B}=\epsilon^{\dot{A}\dot{B}}\sigma^{a\dag}{}_{\dot{B}}{}^{B}$
and $\sigma^{aB\dot{A}}=\epsilon^{BC}\sigma^{a}{}_{C}{}^{\dot{A}}$.
The sigma matrix $\sigma^a$ satisfies a reality condition
\begin{equation}\label{RC4}
\sigma^{a\dag}{}_{\dot{A}}{}^{B}=-\epsilon^{BC}\epsilon_{\dot{A}\dot{B}}\sigma^a{}_{C}{}^{\dot{B}},\quad
{\rm or} \quad\sigma^{a\dag\dot{A}B}=-\sigma^{aB\dot{A}}.
\end{equation}
The antisymmetric matrix $\epsilon_{AB}$ satisfies an important
identity
\begin{equation}
\epsilon_{AB}\epsilon^{CD}=-(\delta_A{}^C\delta_B{}^{D}-\delta_A{}^D\delta_B{}^{C}),
\end{equation}
and $\epsilon_{\dot{A}\dot{B}}$ satisfies a similar identity.

Define \begin{equation}\sigma^{A\dot{B}}\equiv
c_a\sigma^{aA\dot{B}}\quad {\rm and}\quad c_ac^a=1,\end{equation}
where $c_a$ are real coefficients, then the following identity holds
\begin{eqnarray}\label{SU2ID}
\sigma^{A\dot{C}}\sigma^{B\dot{D}}-\sigma^{A\dot{D}}\sigma^{B\dot{C}}=\epsilon^{AB}\epsilon^{\dot{C}\dot{D}}.
\end{eqnarray}
This identity is useful when we construct the ${\cal N}=4$ theory.

Define the parameter for the $\CN=4$ supersymmetry transformations
as $\epsilon^{A\dot{B}}=\ep_a\s^{aA\dot{B}}$. The following
identities are useful in checking the closure of the ${\cal N}=4$
superalgebra:
\begin{eqnarray}\label{SO4ID}
i(\epsilon^{A\dot{C}}_1\epsilon^\dag_{2\dot{C}}{}^B-\epsilon^{A\dot{C}}_2\epsilon^\dag_{1\dot{C}}{}^B)&\equiv&
u^{AB}=u^{BA},\\
i(\epsilon^{\dag\dot{A}C}_1\epsilon_{2C}{}^{\dot{B}}-\epsilon^{\dag\dot{A}C}_2\epsilon_{1C}{}^{\dot{B}})&\equiv&
u^{\dot{A}\dot{B}}=u^{\dot{B}\dot{A}},\nonumber\\
i(\epsilon_{1A}{}^{\dot{A}}\gamma^\mu\epsilon^\dag_{2\dot{A}}{}^B-\epsilon_{2A}{}^{\dot{A}}\gamma^\mu
\epsilon^\dag_{1\dot{A}}{}^B)&=&i\epsilon^{C\dot{C}}_1\gamma^\mu\epsilon_{2C\dot{C}}\delta_A{}^B\equiv
v^\mu\delta_A{}^B,
\nonumber\\
2(\epsilon_{1A\dot{A}}\epsilon^\dag_{2\dot{B}B}-\epsilon_{2A\dot{A}}\epsilon^\dag_{1\dot{B}B})&=&
(\epsilon_{1A}{}^{\dot{C}}\epsilon^\dag_{2\dot{C}B}-\epsilon_{2A}{}^{\dot{C}}\epsilon^\dag_{1\dot{C}B})
\epsilon_{\dot{A}\dot{B}}+(\epsilon^\dag_{1\dot{B}}{}^C\epsilon_{2C\dot{A}}-
\epsilon^\dag_{2\dot{B}}{}^C\epsilon_{1C\dot{A}})\epsilon_{AB},\nonumber\\
i\epsilon_{AB}\epsilon_{\dot{C}\dot{D}}\epsilon^{E\dot{E}}_1\gamma^\mu\epsilon_{2E\dot{E}}
&=&i(\epsilon_{1B\dot{C}}\gamma^\mu\epsilon^\dag_{2\dot{D}A}-
\epsilon_{2B\dot{C}}\gamma^\mu\epsilon^\dag_{1\dot{D}A})-i(\epsilon_{1A\dot{C}}\gamma^\mu\epsilon^\dag_{2\dot{D}B}-
\epsilon_{2A\dot{C}}\gamma^\mu\epsilon^\dag_{1\dot{D}B}).\nonumber
\end{eqnarray}

\subsection{$SO(5)$ Gamma Matrices}\label{SO5}
In this subsection, in order to avoid introducing too many indices
into the theory, we still use the capital letters $A, B, \ldots$ to
label the $Sp(4)$ indices. However, now the index $A$ runs from 1 to
4. (In Sec. \ref{SO4}, the indices $A$ and $\dot{B}$ run from 1
to 2.) We hope this does not cause any confusion.

Since $Sp(4)\cong SO(5)$, it is useful to introduce the $SO(5)$
gamma matrices. We define the $SO(5)$ gamma matrices as
\begin{eqnarray}\label{5Gamma}
\gamma^a_A{}^B=\begin{pmatrix}0&\sigma^a\\\sigma^{a\dag}&0
\end{pmatrix},\quad\quad
\gamma^5_A{}^B=(\gamma^1\gamma^2\gamma^3\gamma^4)_A{}^B ,
\end{eqnarray}
where $\sigma^a$ are defined by (\ref{pulim}). Notice that
$\gamma^m_A{}^B$ ($m=1,\ldots, 5$) are Hermitian, satisfying the
Clifford algebra
\begin{equation}
\gamma^{m}_A{}^C\gamma^{n}_C{}^B+\gamma^{n}_A{}^C\gamma^{m}_C{}^B
=2\delta^{mn}\delta_A{}^B.
\end{equation}
We use an antisymmetric matrix $\omega_{AB}=-\omega^{AB}$ to lower
and raise indices; for instance
\begin{equation}\label{5raise}
\gamma^{mAB}=\omega^{AC}\gamma^m_C{}^B.
\end{equation}
It can be chosen as the charge conjugate matrix:
\begin{equation}
\omega^{AB}=\begin{pmatrix} \epsilon^{AB} & 0 \\
0 & \epsilon^{\dot{A}\dot{B}}
\end{pmatrix}.
\end{equation}
(Recall that $A$ and $\dot{B}$ of the RHS run from 1 to 2.)

By the definition (\ref{5Gamma}) and the convention (\ref{5raise}),
the gamma matrix $\gamma^m$ is antisymmetric and traceless, and
satisfies a reality condition
\begin{eqnarray}
\gamma^{mAB}=-\gamma^{mBA} \quad,\quad \gamma^{m}_A{}^A=0\quad {\rm
and}\quad
\gamma^{m*}_{AB}=\gamma^{mAB}=\omega^{AC}\omega^{BD}\gamma^m_{CD}.
\end{eqnarray}
The $Sp(4)$ generators are defined as
\begin{equation}
\Sigma^{mn}_A{}^B=\frac{1}{4}[\gamma^m, \gamma^n]_A{}^B.
\end{equation}

There is a useful $Sp(4)$ identity
\begin{eqnarray}
\varepsilon^{ABCD}&=&-\omega^{AB}\omega^{CD}
+\omega^{AC}\omega^{BD}-\omega^{AD}\omega^{BC}.
\end{eqnarray}

\section{Verification of $Sp(4)$ Global symmetry of the $\CN=5$ Bosonic
Potential}\label{PotB2} In this section we will prove that the
bosonic potential (\ref{PotB}) has an $Sp(4)$ global symmetry. For
convenience, we cite it here:
\begin{eqnarray}\label{PotBB}
-V&=&\frac{1}{18}f_{IJKO}f^{O}{}_{LMN}(-\omega^{AC}\omega^{BE}\omega^{DF}
+2\omega^{AC}\gamma^{BE}\gamma^{DF}\nonumber\\
&&+2\omega^{DF}\gamma^{AC}\gamma^{BE}-
4\omega^{BE}\gamma^{AC}\gamma^{DF})Z^I_AZ^J_BZ^K_CZ^L_DZ^M_EZ^N_F.
\end{eqnarray}
It can be seen that the first term is manifestly $Sp(4)$ invariant.
So we need only to consider the last three terms. Denote them as
$-V^{\prime}$. For $-V^{\prime}$, the part proportional to
$Z^{(K}_{(C}Z^{L)}_{D)}$ vanishes by the FI (\ref{FFI}), so the
remaining part of $-V^\prime$ is
\begin{eqnarray}\label{PotB4}
-V^{\prime}_{A}&=&\frac{2}{9}( \omega^{AC}\gamma^{BE}\gamma^{DF}-
\omega^{BE}\gamma^{AC}\gamma^{DF})f_{IJKO}f^{O}{}_{LMN}Z^I_AZ^J_BZ^{[K}_{[C}Z^{L]}_{D]}Z^M_EZ^N_F\nonumber\\
&\equiv&\frac{2}{9}(P_1-P_2).
\end{eqnarray}
On the other hand, by using the constraint condition $f_{(IJK)O}=0$
(see (\ref{Constr2})) and the FI (\ref{FFI}), one can rewrite
(\ref{PotB4}) as
\begin{eqnarray}\label{PotB5}
-V^{\prime}_{A}&=&\frac{1}{9}( \omega^{AC}\gamma^{BE}\gamma^{DF}-
\omega^{BE}\gamma^{AC}\gamma^{DF}
+\omega^{CD}\gamma^{AE}\gamma^{BF}-\omega^{BE}\gamma^{CD}\gamma^{AF})
\nonumber\\&&\times
f_{IJKO}f^{O}{}_{LMN}Z^I_AZ^J_BZ^{[K}_{[C}Z^{L]}_{D]}Z^M_EZ^N_F\nonumber\\
&\equiv&\frac{1}{9}(P_1-P_2+P_3-P_4).
\end{eqnarray}
Comparing (\ref{PotB4}) with (\ref{PotB5}) gives
\begin{eqnarray}\label{PotB6}
P_1-P_2=P_3-P_4.
\end{eqnarray}
We observe that $2P_2+P_4$ is an $Sp(4)$ invariant quantity:
\begin{eqnarray}\label{I1}
2P_2+P_4&=&( 2\omega^{BE}\gamma^{A[C}\gamma^{D]F}
+\omega^{BE}\gamma^{CD}\gamma^{AF})
f_{IJKO}f^{O}{}_{LMN}Z^I_AZ^J_BZ^{[K}_{[C}Z^{L]}_{D]}Z^M_EZ^N_F\nonumber\\
&=&\omega^{BE}\varepsilon^{ACDF}f_{IJKO}f^{O}{}_{LMN}Z^I_AZ^J_BZ^{[K}_{[C}Z^{L]}_{D]}Z^M_EZ^N_F\nonumber\\
&\equiv&I.
\end{eqnarray}
In the second line we have used the key identity (\ref{Sp4ID}). By
using the second line of (\ref{Sp4ID}), i.e.
$\varepsilon^{ABCD}=-\omega^{AB}\omega^{CD}
+\omega^{AC}\omega^{BD}-\omega^{AD}\omega^{BC}$, we find that $I$
can be written as
\begin{eqnarray}\label{I2}
-I=\varepsilon_{G}{}^{ACE}\varepsilon^{GBDF}f_{IJKO}f^{O}{}_{LMN}Z^I_AZ^J_BZ^{[K}_{[C}Z^{L]}_{D]}Z^M_EZ^N_F.
\end{eqnarray}
On the other hand, substituting the first line of (\ref{Sp4ID})
($-\varepsilon^{ABCD}
=\gamma^{AC}\gamma^{BD}-\gamma^{BC}\gamma^{AD}+\gamma^{BA}\gamma^{CD}$)
into the RHS of (\ref{I2}), we obtain
\begin{eqnarray}\label{I3}
-I=4P_1-2P_2+P_3.
\end{eqnarray}
Combining (\ref{PotB6}), (\ref{I1}) and (\ref{I3}), we find that
\begin{equation}
P_1-P_2=-\frac{2}{5}I.
\end{equation}
Substituting the above equation into Eq. (\ref{PotB4}), we reach the
desired result:
\begin{equation}
-V^\prime=-\frac{4}{45}I.
\end{equation}
Recall that we denote the last three terms of (\ref{PotBB}) as
$-V^\prime$, so the bosonic potential (\ref{PotBB}) is indeed
$Sp(4)$ invariant. After some work, we reach the final expression
for the bosonic potential (\ref{PotBB}):
\begin{equation}\label{PotBF}
-V=\frac{1}{60}(2f_{IJK}{}^Of_{OLMN}-9f_{KLI}{}^Of_{ONMJ}+2f_{IJL}{}^Of_{OKMN})Z^N_A
Z^{AI}Z^J_BZ^{BK}Z^L_CZ^{CM}.
\end{equation}

\end{document}